\documentclass{ieeeaccess}
\usepackage{cite}
\usepackage{amsmath,amssymb,amsfonts}
\usepackage{algorithmic}
\usepackage{graphicx}
\usepackage{textcomp}
\usepackage{algorithmic}
\usepackage{algorithm}
\usepackage{array}
\usepackage{subcaption}
\usepackage{booktabs}
\usepackage{multirow}
\usepackage{changepage}
\usepackage{url}

\def\BibTeX{{\rm B\kern-.05em{\sc i\kern-.025em b}\kern-.08em
    T\kern-.1667em\lower.7ex\hbox{E}\kern-.125emX}}

\begin{document}

\newcommand{\norm}[1]{\left\lVert#1\right\rVert}

\history{Received 30 April 2024, accepted 26 May 2024, date of publication 3 June 2024, date of current version 19 June 2024.}
\doi{10.1109/ACCESS.2024.3409067
}

\title{Privacy-oriented Manipulation of Speaker Representations}
\author{\uppercase{Francisco Teixeira}\authorrefmark{1},
\uppercase{Alberto Abad}\authorrefmark{1}, \IEEEmembership{Senior Member, IEEE}, \uppercase{Bhiksha Raj}\authorrefmark{2,3}, \IEEEmembership{Fellow, IEEE}, and \uppercase{Isabel Trancoso}\authorrefmark{1},
\IEEEmembership{Life Fellow, IEEE}
}

\address[1]{INESC-ID/Instituto Superior Técnico, University of Lisbon, Lisbon, Portugal}
\address[2]{LTI, Carnegie Mellon University, Pittsburgh, Pennsylvania, USA}
\address[3]{Mohammed bin Zayed University of AI, Abu Dhabi, UAE}

\tfootnote{This work was supported in part by Portuguese National Funds through Fundação para a Ciência e a Tecnologia with reference 10.54499/UIDB/50021/2020, and in part by the Recovery and Resilience Plan and Next Generation EU European Funds under Grant C644865762-00000008 Accelerat.AI.}

\markboth
{F. Teixeira \headeretal: Privacy-oriented manipulation of speaker representations}
{F. Teixeira \headeretal: Privacy-oriented manipulation of speaker representations}

\corresp{Corresponding author: Francisco Teixeira (e-mail: francisco.s.teixeira@inesc-id.pt).}

\begin{abstract}
Speaker embeddings are ubiquitous, with applications ranging from speaker recognition and diarization to speech synthesis and voice anonymization. The amount of information held by these embeddings lends them versatility but also raises privacy concerns. 
Speaker embeddings have been shown to contain sensitive information, including the speaker’s age, sex, health state and more – in other words,  information that speakers may want to keep private, especially when it is not required for the target task. 
In this work, we propose a method for removing and manipulating private attribute information in speaker representations that leverages a Vector-Quantized Variational Autoencoder architecture combined with an adversarial classifier and a novel mutual information loss. We validate our model on two attributes, sex and age, and perform experiments to remove or manipulate this information using ignorant and informed attackers. The model is tested with in-domain and out-of-domain data to assess its robustness, and the resulting speaker representations are used in a speaker verification scenario to validate their utility. Our results show that our model obtains a strong trade-off between utility and privacy, achieving age and sex classification results near chance level for both attackers and yielding little impact on speaker verification performance.

\end{abstract}


\titlepgskip=-21pt

\maketitle

\section{Introduction}
\label{sec:intro}
\PARstart{S}{peaker} representations, or embeddings -- vector representations that model speakers' voices -- are a key component in speech technologies.
Originally developed for speaker recognition \cite{gmmsupervectors, i-vectors, x-vectors}, i.e., the task of identifying or verifying the identity of a speaker, speaker embeddings are applied to a multitude of tasks that extend far beyond their original purpose.

Traditional speaker embedding extractor systems were built to model how speech was produced by a speaker, relying on generative models such as Gaussian Mixture Model - Universal Background Models (GMM-UBM) \cite{gmm-ubm}, Gaussian Mixture Model (GMM) Supervectors \cite{gmm-super} and \textit{i-vectors} \cite{i-vectors}. Modern neural speaker embedding extractor systems such as \textit{d-vectors} \cite{variani2014deep} and \textit{x-vectors} \cite{x-vectors,desplanques2020ecapa,zhang2022mfa} on the other hand, model the differences between speakers by relying on latent representations. These are extracted from intermediate layers of deep neural network models which are trained to classify large sets of speakers, hence being considered discriminative systems.

Applications of neural speaker embeddings~\cite{desplanques2020ecapa, zhang2022mfa} range from speaker diarization~\cite{landini2022bayesian}, to text-to-speech synthesis~\cite{cooper2020zero}, voice anonymization \cite{tomashenko2020introducing}, and even detection of speech-affecting diseases~\cite{perero2019modeling,egas2022automatic}.

This versatility is a testament to the wealth of information that is encoded by neural speaker embeddings, including (i) linguistic information~\cite{raj2019probing, quintas2022automatic}; (ii) paralinguistic information~\cite{laver1994principles}, i.e., non-linguistic, but communicative information, such as affective, attitudinal and emotional information~\cite{pappagari2020x,juliao2020}; and (iii) extra-linguistic information~\cite{laver1994principles}, i.e. non-communicative information about the speaker that is carried by the speech signal, such as the speaker's age and sex~\cite{kwasny2020joint}, accent~\cite{raj2019probing}, as well as the speaker's health state (i.e., the presence of speech-affecting diseases such as Parkinson's disease or Obstructive Sleep Apnea, among others)~\cite{Parkinson-x-vectors, perero2019modeling}.
However, whereas this information renders speaker representations particularly useful, it also raises questions of privacy and even adherence to data protection regulations when speaker representations are processed outside users' devices.

Under the definitions introduced by the European Union's General Data Protection Regulation (GDPR) \cite{gdpr}, and similar data protection regulations~\cite{ccpa}, speech data and representations derived from it may be considered biometric data, and, by extent, sensitive personal data \cite{nautsch2019gdpr, nautsch2019preserving}.
As such, remote speech data processing should adhere to the \textit{privacy-by-design} and data protection principles enshrined by Article 25 of the GDPR~\cite{gdpr}.

Such legal -- and ethical -- concerns have motivated a significant number of studies on privacy-preserving remote speech data processing.
Two main types of approaches have been considered for the problem of privacy in remote speech processing: cryptographic protocols and speech manipulation methods. 

%%% Existing solutions
Cryptographic techniques such as Homomorphic Encryption \cite{cheon2017homomorphic} or Secure Multiparty Computation protocols \cite{lindell2020secure} allow two or more parties to compute functions over their data securely. These protocols are applied collaboratively between different parties (e.g., client and remote service provider), with each operation performed over the parties' data being replaced by its cryptographic counterpart.
Such techniques provide guarantees of confidentiality and security and can be applied such that only users can see the result of the operations performed over their data.

Recent years have seen increasingly complex systems being implemented with these techniques \cite{nautsch2018homomorphic, treiber2019privacy, wang2020novel, teixeira2022towards, teixeira2023privacy}; however, the computational and communication costs of the resulting methods are still high, and are limited by the state-of-the-art of the underlying cryptographic constructions. Moreover, the computational performance of these methods depends on the complexity of the target task, making them difficult to apply to state-of-the-art systems that leverage machine learning models that require billions of operations. % giga flops - 10^9

Privacy-oriented speech manipulation methods have a different goal. Instead of providing confidentiality during the computation, these methods are applied before the data is processed and aim to remove or sanitise information that is considered private and not relevant to the target task~\cite{tomashenko2020introducing, aloufi2019emotionfiltering, noe2021adversarial}. This allows for a conscious trade-off between the information that is disclosed and the information that should remain hidden, or in other words, a trade-off between privacy and utility. 
These solutions are also more user-centred, as the privatisation process may be applied directly in the users' devices \cite{aloufi2019emotionfiltering, wu2021privacy}. 

Speech manipulation methods also go in line with the \textit{data minimisation} principle mentioned in Article 25 of the GDPR and defined in Article 5 of the GDPR, whereby personal data should be ``adequate, relevant and limited to what is necessary in relation to the purposes for which they are processed''~\cite{gdpr}. 

These methods have the advantage of being independent of the downstream task's complexity, though not necessarily of the task itself. This is an advantage over cryptographic protocols as it allows the downstream adoption of arbitrarily complex state-of-the-art methods.
However, unlike cryptographic constructions, this family of methods does not provide any formal privacy guarantees. This means that the evaluation of these methods, which is usually done empirically, needs to be thorough and well-designed to adequately support privacy claims.

Privacy-oriented speech manipulation methods follow three main trends. 
The first is voice anonymisation \cite{tomashenko2020introducing}, where the goal is to modify the speech signal to hide the identity of the true speaker but keep linguistic and paralinguistic content intact, such that the speech signal is considered anonymised under the GDPR, allowing its storage and use in the training of speech-based machine learning applications, or even in remote inference scenarios, where only linguistic or paralinguistic content are necessary for the task at hand.
The second trend is privacy-oriented feature extraction \cite{nelus2019privacy, wang2023non}, where the goal is to obtain feature vectors from which all of the information that is not related to the target task is removed and where particular focus is given to the removal of speaker-identity-related information.
The third trend consists of attribute disentanglement, manipulation, or removal methods. This is a more fine-grained approach that aims to remove specific speaker traits that are considered sensitive from the speech signal, or a representation thereof, while keeping the remaining information intact~\cite{aloufi2019emotionfiltering, noe2021adversarial, perero2022xvector}.

In this work, we focus on the third trend and propose a method for attribute manipulation and removal in speaker embeddings. 
As mentioned at the beginning of this section, neural speaker representations have a very large number of applications. Consequently, modifying these representations to promote privacy will indirectly lend a level of privacy to downstream applications.
For instance, removing demographic attributes from speech (or speech representations) can potentially avoid negative biases or even discrimination on the part of the service provider.
Moreover, as shown by \cite{perero2022xvector, noe2023hiding}, privatised speaker representations can be used to perform voice anonymisation to a certain extent.

Notwithstanding other possible applications, the primary purpose of speaker embeddings is to perform Automatic Speaker Verification (ASV), the process of verifying an individual's identity through their voice -- a process which is performed mainly in remote settings.
Privatised representations that hide sensitive speaker attributes will directly prevent speaker verification vendors (remote servers) from inferring sensitive information, again providing a level of privacy to this task \cite{noe2021adversarial, noe2022bridge, chouchane2023differentially}. 
Given that ASV is the main and original application of speaker embeddings, and that measuring ASV performance using privatised vectors provides an estimate of how much the vectors' original (non-private) content was changed, we consider ASV as both our target task and measure of utility.

The contributions of this work are summarised below:
\begin{itemize}
    \item We propose a new method for the privacy-oriented removal and manipulation of age and sex information in speaker representations. To the best of our knowledge, this work is the first to consider the removal of age information from speaker representations.
    \item Our method is based on a combination of a Vector Quantised Variational Autoencoder (VQ-VAE), an adversarial classifier and a novel mutual information loss.
    \item For each attribute we evaluate our method with two competing aspects: privacy and utility. 
    \begin{itemize}
        \item[--] Privacy is assessed using as a proxy the attribute classification performance of two types of attackers, an ignorant attacker and an informed attacker.
        \item[--] Utility is evaluated in terms of ASV performance.
        \item[--] We perform an ablation study to assess the privacy and utility contributions of each component of our method.   
        \item[--] We evaluate the attribute manipulation performance of the proposed methods, to understand whether they are versatile enough to be applied in tasks that are not related to privacy.
    \end{itemize}
    \item For the sex attribute:
    \begin{itemize}
         \item[--] We evaluate our method through its performance on out-of-domain data, to assess its transferability to new domains. 
    \end{itemize}
    \item Overall, our results show that the proposed mutual information loss improves both privacy and utility when combined with the adversarial classifier, with their combination being able to reach near chance-level classification for both attributes and types of attackers. The proposed model is also shown to transfer to new domains and to be able to successfully manipulate attribute information within the speaker representations.
\end{itemize}

The remainder of this paper is organised as follows: Section \ref{sec:related_work} provides an overview of the relevant literature; in Section \ref{sec:prelim}, we formally describe the problem at hand; Section \ref{sec:method} presents the proposed method and each of its components; Section \ref{sec:exp} details the experiments that were conducted along with the corresponding datasets and parameters; in Section \ref{sec:results}, we present and discuss our results, and in Section \ref{sec:conclusions}, we provide closing statements and propose topics for future work.

\section{Related Work}
\label{sec:related_work}
%%%%%%%%%% Related Work %%%%%%%%%%%%

Modifying or suppressing speaker attributes within the speech signal, or representations thereof, is a growing area of research. Several studies do so to ensure that classifiers are invariant with regard to certain speaker traits \cite{perero2019modeling,luu2022investigating,janbakhshi2022adversarial}, or to create control mechanisms for speech synthesis and voice conversion algorithms \cite{benaroya2023manipulating}. In addition to this, and more relevant to the present work, privacy-related approaches have also seen a surge in recent years.

An early example of attribute suppression for privacy is the work of Aloufi et al. \cite{aloufi2019emotionfiltering}, where the authors apply a CycleGAN to convert emotional speech to neutral speech as a way to remove sensitive, emotional information from the speech signal.
In \cite{aloufi2020privacy, aloufi2023paralinguistic}, the same authors proposed two methods to protect the privacy of speaker identity, emotional content, sex, and accent/language information. 
This is done to protect the user's privacy for Automatic Speech Recognition (ASR). 
The methods are based on encoder-decoder architectures, whose encoders comprise two branches, one encoding linguistic information and another encoding speaker or paralinguistic information. 
By selecting the branches that are fed to the decoder, the authors can select the information present in the output signal. 
In \cite{aloufi2023paralinguistic}, the authors evaluate their model in terms of efficiency to assess its usability in the context of mobile computing. 

Jaiswal et al. \cite{jaiswal2020privacy} develop a neural network for emotion classification using speech and text data. 
This network includes an adversarial classifier with a Gradient Reversal Layer (GRL)~\cite{ganin2015unsupervised} that promotes the learning of latent representations that are invariant to sex, making them private in relation to this attribute. 
The authors show that their method has little impact on emotion classification performance while improving privacy protection, to varying degrees, with respect to sex information.
The authors also study how their sex-invariant representations affect an attacker's ability to perform membership inference (i.e., classify whether a sample was seen or not during the model's training).

Ericsson et al. \cite{ericsson2020adversarial} proposed a model to remove sex information from speech and validate their model for spoken digit classification. 
Similarly to \cite{aloufi2020privacy, aloufi2023paralinguistic}, this method is based on an encoder-decoder network, where the encoder acts as a filter to the sensitive attribute, and the decoder takes this sanitised representation and reconstructs the speech signal using a fake, externally provided attribute. To promote the removal of sex information, the filter is trained adversarially against the attribute classifier.

Stoidis and Cavallaro \cite{stoidis2021protecting} focused on disentangling and manipulating sex and speaker identity from the speech signal for privacy using a VQ-VAE and evaluated the utility of their method through ASR performance. 
Later, the same authors developed a method based on their prior work and the work of Ericsson et al. \cite{ericsson2020adversarial}, to generate gender-ambiguous voices (i.e. voices that are not strongly related to any gender) for ASR~\cite{stoidis2022generating}. 

Wu et al. \cite{wu2021privacy} explore and compare multiple methods to remove sex and accent from speech, including pitch standardisation, a Variational Autoencoder (VAE), and a version of the same VAE combined with a Generative Adversarial Network for improved speech reconstruction quality. The VAE was found to be the best-performing model for privacy protection.

Differently, Bemmel et al. \cite{bemmel2023beyond} study the protection provided by adversarial examples created against sex classification neural networks. The authors show that combining a simple Support Vector Machine with knowledge-based features for sex classification is sufficient to overcome the adversarial perturbation and successfully classify sex. The authors also propose the use of different vocal adaptations (e.g. whispering, monotonality, high pitch) as protection against sex classifiers that use knowledge-based features.

Whereas the approaches above have focused on removing information from or hiding information contained in the speech signal itself, other works have instead focused on removing information from speaker representations or knowledge-based feature vectors.

In No\'e et al. \cite{noe2021adversarial}, this is done through the use of an Autoencoder (AE) trained adversarially against a sex classifier where, in the same fashion as \cite{ericsson2020adversarial}, the decoding part of the network is conditioned on an externally provided attribute.

Similarly, Ali et al. \cite{ali2021privacy} propose the use of an autoencoder architecture with an adversarial branch, using a Gradient Reversal Layer, so that the encoder learns to remove sex, language, and speaker information from a set of speech features while keeping the remaining content intact. This approach is then applied to remote emotion recognition.

In \cite{noe2022bridge}, the same authors of \cite{noe2021adversarial} propose the use of a Normalising Flow-based architecture that disentangles sex information and aggregates it in a single component in a latent representation of the speaker embedding. To remove sex information, the component in the latent representation is set to zero, and the vector is reconstructed. In the same paper, \cite{noe2022bridge}, the authors also argue that to assess how well an attribute is removed, attacker classifiers should be trained over protected representations.

Feng and Narayanan \cite{feng2021privacy}, in a similar line to that of \cite{aloufi2019emotionfiltering}, develop a model to transform the emotional content of a knowledge-base feature vector into a neutral emotion, in case the corresponding emotion is deemed sensitive (e.g. anger). The resulting transformed vector is then used to infer non-sensitive emotions (e.g. sadness). An adversarial classifier is further added to remove sex information from the feature vector. 
Later, within the same emotion recognition context, Feng et al.~\cite{feng2022enhancing} used a multi-objective mutual information-based feature selection approach, to select the set of features that were most relevant for emotion classification and least informative regarding speaker sex. This approach also included the addition of Gaussian noise tailored to the masking of sex information, in addition to an adversarial classifier that was added to remove sex information from the resulting features.

Similar to \cite{noe2021adversarial, noe2022bridge}, Perero-Codosero et al. \cite{perero2022xvector} propose the use of an adversarial autoencoder, based on their prior work \cite{perero2019modeling}, to remove speaker identity, sex and accent information from speaker representations. To remove each of these, an adversarial classifier with a GRL is added and applied over the latent representations of the autoencoder. The privatised speaker representations are subsequently used as part of a voice anonymisation framework. 

Recently, Chouchane et al. \cite{chouchane2023differentially}, basing their approach on the work of No\'e et al. \cite{noe2021adversarial}, proposed a method where differentially private noise is added to an autoencoder's latent representation, to remove sex information from a speaker representation. The authors show that, by controlling the level of noise, they can achieve different trade-offs between privacy and utility (i.e. speaker verification performance).

As mentioned in Section~\ref{sec:intro}, one of the main trends of privacy-oriented speech manipulation is privacy-aware feature extraction.
The main goal in this research line is to remove all of the information that is not necessary to the target task, while simultaneously optimising the representation for the target task. Although this goal differs from ours, it is worth mentioning some works related to this trend, as they share many of the methods used for attribute suppression.

For instance, Nelus and Martin~\cite{nelus2018adversarial} proposed an adversarial training architecture to remove speaker information from a feature representation used to classify speaker sex. In a later work~\cite{nelus2019privacy}, the same authors apply the concept of a variational information bottleneck and minimise the mutual information between the input and output representations of a neural network trained for sex classification. This is done to minimise the amount of information contained by the feature representation that is not relevant to the target task. It is then shown that this reduces the amount of information related to speaker identity. Building on their two prior works, in ~\cite{nelus2019siamese}, Nelus and Martin train a neural network for sex classification using a Siamese architecture trained with a contrastive loss, to bring feature vectors that belong to speakers from the same sex closer together, and vice-versa. The authors show that the latter approach obtains improved results both in terms of utility and speaker privacy when compared to the two previous works.

Similarly, the work of Wang et al.~\cite{wang2022optin} focuses on the removal of all target-task irrelevant information, as opposed to the removal of selected attributes. To this end, the authors leverage a CycleGAN ``obfuscator'', trained to minimise a target task loss (e.g. sex or speaker classification), while simultaneously being trained adversarially against a ``deobfuscator'' that attempts to reconstruct the true signal from the obfuscated signal. This combination is then expected to elicit the model to remove all information that is unnecessary to the target task.

The works of Ravi et al.~\cite{ravi2022steo,ravi2024enhancing} and Wang et al.~\cite{wang2023non} focus on the development of privacy-aware feature extraction methods for the classification of depression, while removing all non-depression-related speaker information, using adversarial training. 
Whereas Ravi et al. \cite{ravi2022steo} focus solely on adversarial training, in \cite{ravi2024enhancing} the authors expand their previous work, testing several models and different adversarial loss functions. Although the three works leverage a GRL, Wang et al.~\cite{wang2023non} propose a variation of the work of \cite{ravi2022steo} by assigning different gradient weights to different layers, which is shown to improve the trade-off between target task performance and privacy.

Though not related to privacy, the works of Janbakhshi and Kodrasi~\cite{janbakhshi2022adversarial}, Mun et al.~\cite{mun2022disentangled}, and Li et al.~\cite{li2023mutual} are also worth mentioning, due to their use of mutual information-based losses for information disentanglement.
Specifically, Janbakhshi and Kodrasi~\cite{janbakhshi2022adversarial} propose a method for the detection of dysarthric speech that aims to be invariant with respect to speaker information. To this end, the authors use an AE architecture, trained to reconstruct the input signal, using two branches, one to encode task-related information, and a second to encode speaker information. Both encoders are trained to classify the information they are meant to encode. To promote information independence between the two branches, the authors add a mutual information minimisation loss which is based on the CLUB mutual information upper bound~\cite{cheng2020club}. Similar approaches have been used by Mun et al.~\cite{mun2022disentangled} and Li et al.~\cite{li2023mutual} to disentangle speaker information and domain conditions for improved domain generalisation in speaker recognition tasks.

It is also important to note that there are template protection mechanisms that can perform privacy-preserving enrolment and authentication in ASV, concealing all of the user's information~\cite{smh,mtibaa2018cancelable,mtibaa2022towards}. 
These mechanisms correspond to transformations of the input, such that the original values cannot be recovered from the transformed ones. This makes these schemes secure, as any party can hold the transformed vector without being able to learn any information about it. Moreover, vectors transformed in the same way (i.e., using the same secret key) can be meaningfully compared.
Although such schemes are important to biometric verification, they are not directly applicable to tasks other than verification, retrieval or clustering. In contrast, the method developed in this work extends to any downstream task, even though it does not provide confidentiality.

\section{Formal problem definition}
\label{sec:prelim}

%%%%%%%%%% Preliminaries %%%%%%%%%%%%

As mentioned in Section \ref{sec:intro}, in this work, we consider a remote Automatic Speaker Verification scenario, where a user wants to be able to authenticate %themselves 
through a remote ASV service provider (or vendor).
To do so, the user first needs to enrol into the system by sending a speaker embedding to be used as a template. Later, for authentication, the same user generates a new embedding of their voice and sends it to the vendor so that the vendor can compare it to the stored template.

In this scenario, we assume that the speaker representation is extracted on the user's device whereas verification is performed remotely.
We also assume that the user does not fully trust the service provider with their information and wants to hide sensitive attributes contained in the speaker representations, such that the service provider or any other entity that can obtain the user's speaker representation (e.g., via a data breach, or directly shared by the ASV vendor), is not able to infer the sensitive information from it.

ASV was chosen as our target task as it represents a simple setting where we can test the utility and privacy of the transformed speaker representations. 

The scenario described above can be simplified as an adversarial game, where we have a user trying to protect sensitive attribute information about themselves and an attacker who wants to obtain this information.
As such, we want to develop a method of hiding a sensitive attribute from a speaker representation so that an attacker cannot obtain this attribute just by observing the transformed representation. This method should be applied in the user's device after the speaker representation has been extracted.

For a given input speaker embedding $x$ with private attribute $y_a$, \textit{discrete or continuous}, coming from a dataset $\mathcal{D}$, our goal is to learn a function $F_a$ that removes attribute information $y_a$. Moreover, for versatility, we want our method to not only remove attribute information but also to be able to manipulate it.
As such, we want to develop a function $F_a$ that removes $y_a$ and replaces it with external information $\hat{y}_a$:

\begin{equation}
    \hat{x} = F_a(x | \hat{y_a})
    \label{eq:f_a}
\end{equation}

To ensure the attacker is not able to learn anything about the attribute, we should select $\hat{y_a}$ such that it provides the least amount of information -- e.g., using the expected value of $y_a$.
Nevertheless, defining our model as dependent on the conditioning of the decoder allows us to choose the best strategy to undermine a possible attacker.

To ensure utility, we also want $F_a$ to guarantee the same discriminability shown by the original vectors. In other words, transformed vectors that belong to different speakers should be far apart, whereas those that belong to the same speaker should be as close as possible. To measure this, we can compute the distance of the same- and different-speaker pairs of vectors after transformation and measure how discriminative this distance is, concerning speaker identity.

To measure the level of privacy provided by $F_a$, we need to assess how well an attacker can recover the original attribute $y_a$. 
However, an attacker can take different forms. Here, we consider two types of attackers with different levels of knowledge about the protection mechanism: an \textit{ignorant attacker} and an \textit{informed attacker}.

We assume that the weakest possible attacker, the \textit{ignorant attacker}, will try to infer the original attribute directly, having no knowledge of the privatisation mechanism.
We assume that an \textit{ignorant attacker}, will hold an attribute classifier $C_{\mathcal{A}}$, trained on a dataset $\mathcal{D} = \{(x_1, y_1), (x_2, y_2), ... (x_n, y_n)\}$ of non-transformed data, with probability $\mathbb{P}(C_{\mathcal{A}}(x)=y_a)$ as close to 1 as possible.

In the case of classification, to guarantee privacy with regard to $y_a$, the following should hold for any pair $(x, y_a)$:
\begin{equation}
    \mathbb{P}(C_{\mathcal{A}}(F_a(x | \hat{y}_a)) = y_a) = \frac{1}{n_a},
    \label{eq:privacy}
\end{equation}

\noindent with $n_a$ as the number of classes of attribute $a$.

To encompass the possibility of $F_a$ allowing the manipulation of the attribute $y_a$ within the speaker embedding, we also want that $\mathbb{P}(C_{\mathcal{A}}(\hat{x}) = \hat{y_a})$ be as high as possible. This means that an attacker holding any classifier trained on non-transformed data should not be able to obtain any information about attribute $y_a$ by observing $\hat{x}$ unless the fake attribute $\hat{y}_a$ is the same as the true attribute~$y_a$:
\begin{equation}
    C_{\mathcal{A}}(F_a(x | \hat{y}_a)) = y_a \leftrightarrow y_a=\hat{y}_a.
    \label{eq:manipulation}
\end{equation}

Still, to ensure that the information is fully protected, we need to account for the possibility of an attacker being aware of the transformation that was applied to the speaker representation. As such, we consider as a stronger attacker, the \textit{informed attacker}. This attacker not only knows that a privacy transformation was put in place but is also able to apply this transformation to its data, for which the true labels are known, and train a classifier using the privatised representations.
In a way, this attacker will develop a classifier to try to infer the sensitive attribute, using the residual information that is still encoded by the privatised representations.
We assume that this attacker will hold an attribute classifier $\hat{C}_{\mathcal{A}}$, trained on a dataset $\mathcal{\hat{D}} = \{(\hat{x}_1, y_1), (\hat{x}_2, y_2), ... (\hat{x}_n, y_n)\}$ of data transformed as $\hat{x} = F_a(x | \hat{y}_a)$.
In this situation, our goal is that the attribute classifier trained by the \textit{informed attacker} is not able to generalise beyond the training data such that, for unseen data, $\mathbb{P}(\hat{C}_{\mathcal{A}}(\hat{x}) = y_a) = \frac{1}{n_a}$.

To summarise the above, the goal of this work is to develop a method that achieves the following under the two attack scenarios:
\begin{itemize}
    \item Allows the suppression of attribute information from speaker representations and enforces privacy regarding this subset of information (cf. eq. \eqref{eq:privacy});
    
    \item It not only removes attribute information but manipulates it within the speaker embedding (cf. eq. \eqref{eq:manipulation});
    
    \item Keeps the utility of the transformed vectors for speaker verification.
\end{itemize}

\section{Method}
\label{sec:method}

\begin{figure*}
    \centering
    \includegraphics[width=0.8\textwidth]{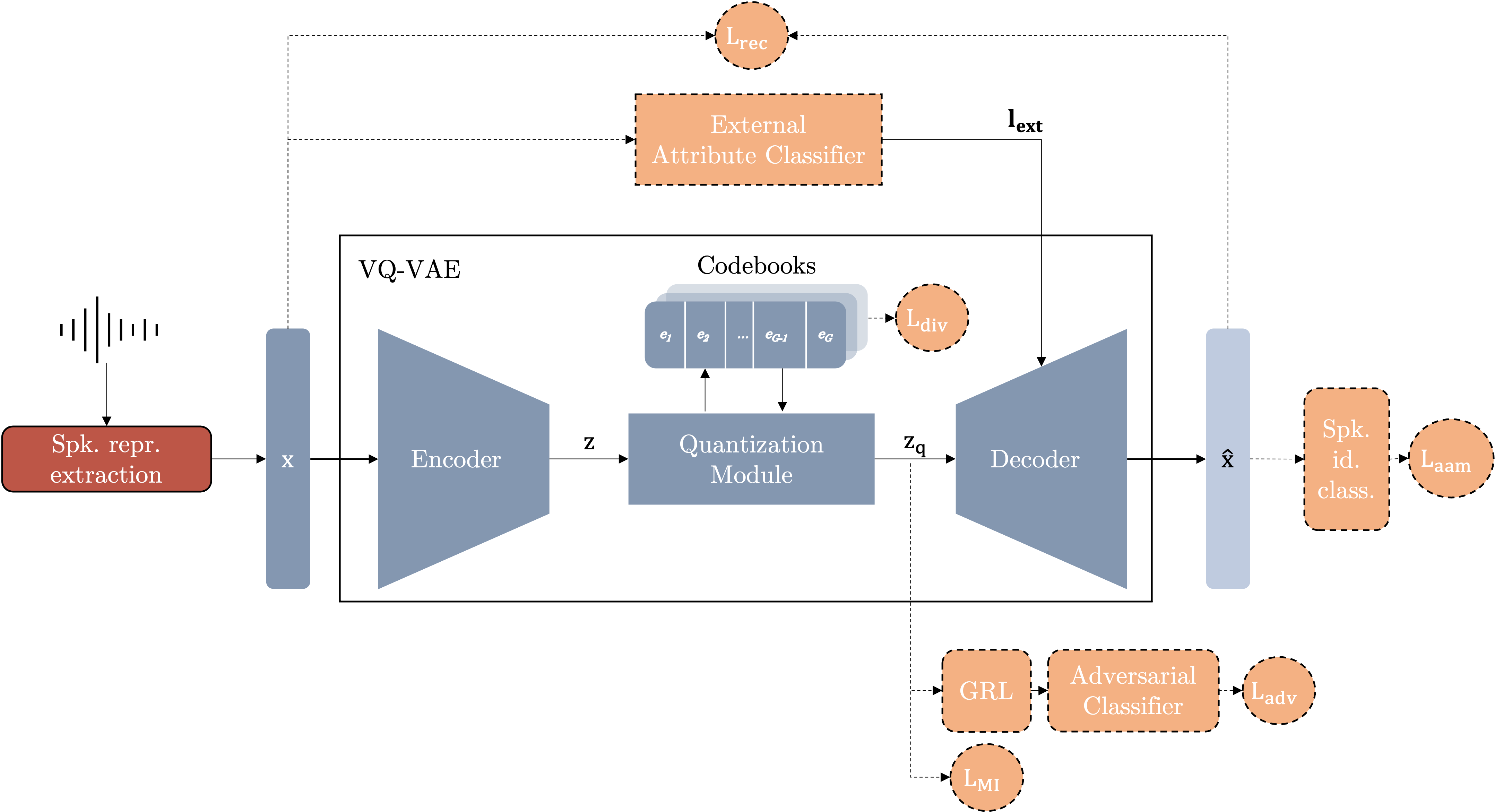}
    \caption{Block diagram of the proposed method. Dashed boxes and lines represent components that are only necessary during training and that are dropped at inference time.}
    \label{fig:method}
\end{figure*}

%%%%%%%%%% Method %%%%%%%%%%%%

To achieve the objectives summarised in the previous section, we propose a combination of five components: a Vector-Quantized Variational AutoEncoder (VQ-VAE); an external speaker identification classifier; an external attribute classifier $C_{ext}$; an adversarial attribute classifier, $C_{adv}$; and a Mutual Information (MI) loss $L_{\text{MI}}$.
In the remainder of this section, we will detail each of these components and their role in removing information from speaker representations.

\subsection{Vector-Quantized Variational Autoencoder}
\label{sec:vqvae}

The main basis of our method is a Vector Quantised Variational Autoencoder (VQ-VAE). VQ-VAEs have been shown to perform well for several speech tasks \cite{van2017neural, baevski2019vq, baevski2020wav2vec2}, revealing a solid capability for information disentanglement~\cite{van2017neural,chorowski2019unsupervised,wu2020one}.
In this section, we briefly introduce the concept of VQ-VAEs and detail the importance of this model in our overall method.

Variational Autoencoders (VAEs)~\cite{kingma2014vae} are a family of generative models that have been widely used for synthetic data generation, representation learning and disentanglement. VAEs follow a general autoencoder architecture, being composed of an \textit{encoder} and a \textit{decoder}. Specifically, the encoder creates a latent representation from the input, while the decoder uses this representation to reconstruct the input. During training, the encoder learns to map the input to the parameters of a prior distribution -- usually, a normal distribution parameterised by a mean vector and a covariance matrix -- while the decoder learns to reconstruct the input by sampling from this distribution. This, together with its specific loss function, regularises the latent space, imposing a structure on the model's latent representations. This property makes it possible to use the decoder as a generator by sampling from the latent space. In addition, the structured latent space will be composed of independent, or disentangled, factors, allowing for an easier manipulation of the input signal when represented in this form.

However, VAEs have been shown to suffer from poor reconstruction quality, and, when combined with more powerful decoders to improve quality, VAEs often suffer from posterior collapse~\cite{van2017neural}, i.e., the decoder ignores the latent representation when producing the output, thus ignoring most, if not all, of the information coming from the input.

To address these issues, van den Oord et al.~\cite{van2017neural} proposed a vector quantised version of VAEs (VQ-VAE). In this version, instead of being modelled by a continuous prior distribution, the latent space is modelled by a learnable set of discrete codes. To perform inference, this set of codes, the \textit{codebook}, is indexed by the output of the encoder, which selects the sub-set of codes that best models the input. The decoder then takes this sub-set of codes and reconstructs the input.

This poses several advantages over the original VAE, namely avoiding the problem of posterior collapse, by having a function of the input select the codes that best model it, and improves reconstruction quality, by the fact that the latent space is no longer static, being trainable, and thus more adjusted to the training data distribution. Moreover, the discrete nature of the codebook also helps in the disentanglement of information, as each entry in the codebook will correspond to an aspect of the input signal.

When considering our target task, the removal and manipulation of information within a speaker representation for privacy, VQ-VAEs appear as an attractive solution.
This comes from the fact that all of the information that is necessary to reconstruct the input signal is obtained from the quantization module and that this information is inherently disentangled, making it easier to manipulate or remove.

Formally, a VQ-VAE is defined as follows~\cite{van2017neural}: assume we have an encoder $E: \mathbb{R}^n \rightarrow \mathbb{R}^h$, a decoder $D: \mathbb{R}^f \rightarrow \mathbb{R}^n$ and a quantization module $Q: \mathbb{R}^h \rightarrow \mathbb{R}^q$. For an input vector (in our case a speaker embedding) $x \in \mathbb{R}^n$, we start by feeding it through the encoder $E$ to obtain a latent representation $\mathbf{z} \in \mathbb{R}^h$; this vector is passed through the quantization module, where we obtain the quantized representation $\mathbf{z}_q \in \mathbb{R}^q$; $\mathbf{z}_q$ is in turn fed to decoder $D$, such that the original input is reconstructed.

Our setting differs from a regular VQ-VAE because we want the output to differ from the input. However, we do not have access to embeddings of the same speaker presenting different versions of each attribute. 
As such, to be able to train the VQ-VAE and promote attribute disentanglement, we turn to the solution of No\'{e} et al.~\cite{noe2021adversarial} and condition the decoder with the output of an external pre-trained attribute classifier.

Specifically, we take the output logits $l_{ext}$ of an external classifier $C_{ext}: \mathbb{R}^n \rightarrow \mathbb{R}^{c_{\text{attr}}}$ -- where $c_{\text{attr}}$ corresponds to the number of classes\footnote{$c_{\text{attr}}$ = 1 for regression tasks.} -- obtained for the original input, to which we apply a linear transformation $h_{attr}: \mathbb{R}^{c_{\text{attr}}} \rightarrow \mathbb{R}^w$ and concatenate this representation with the output of the quantization module, $\mathbf{z}_q$, obtaining:
\begin{equation}
    \mathbf{\hat{z}}_q = [ \mathbf{z}_q\ | \ h_{attr}(l_{ext}) ],
\end{equation} 
\noindent where $|$ represents the concatenation operator; $\mathbf{\hat{z}}_q$ is then feed as input to the decoder $D$.

This enables the VQ-VAE to reconstruct the original input signal during training while also allowing us to manipulate the attribute information at test time by changing the values used to condition the decoder. Moreover, it also provides an implicit level of disentanglement, as the decoder will not require as much information about the attribute from the latent representation, since it has direct access to it from the conditioning logits.

\subsubsection{Quantization Module}
Our implementation of the quantization module of the VQ-VAE corresponds to the product quantization approach of Baevski et al. \cite{baevski2020wav2vec2, jegou2010product}.
In \cite{baevski2020wav2vec2}, the quantization module is defined as a tensor $Q \in \mathbb{R}^{G\times V\times{e/G}}$, with $G$ being the number of codebooks, and $V$ the number of codewords $v \in \mathbb{R}^{e/G}$ within each codebook. 
To quantize a latent vector $\mathbf{z} = E(x)$, we select an entry $v$ from the $V$ entries of each codebook $G$ to obtain a set of codewords $v_1, ..., v_G$.
To this end, first, a linear transformation is applied $\mathbb{R}^{h}\rightarrow\mathbb{R}^{G*V}$, to obtain $\mathbf{\hat{z}} \in \mathbb{R}^{G*V}$, after which $\mathbf{\hat{z}}$ is reshaped to $\mathbb{R}^{G\times V}$, giving us G logit vectors $l_g \in \mathbb{R}^V$ (one logit per codeword per codebook).
To choose entries $v$ at inference time, the largest index $i$ of each $l_g$ is selected.
During training, to ensure the selection is fully differentiable, a straight-through estimator of the Gumbel-Softmax is used~\cite{baevski2019vq, baevski2020wav2vec2, jang2016categorical}:

\begin{equation}
    p_{g,v}= \frac{\text{exp}{(l_{g,v} + \eta_v)/\tau}}{\sum_{k=1}^{V} \text{exp}{(l_{g,k} + \eta_k)/\tau}},
    \label{eq:gumbel_softmax}
\end{equation} 

where each $p_{g,v}$ corresponds to the probability of selecting entry $v$ of codebook $g$; $\eta_v = -\text{log}(-\text{log}(u_v))$, with $u_v$ uniformly sampled from $\mathcal{U}(0, 1)$; and $\tau$ is a non-negative temperature. 
During the forward pass, the codeword is selected by index $i = \text{argmax}_j p_{g,j}$, whereas in the backward pass, the true gradient of eq. \eqref{eq:gumbel_softmax} is used.
After $v_1, ..., v_G$ have been selected, a final linear transformation is applied, $\mathbb{R}^{e} \rightarrow \mathbb{R}^q$, to obtain $\mathbf{z}_q \in \mathbb{R}^{q}$. 

\subsubsection{Training losses}
The VQ-VAE is trained with several losses. The first loss we consider is the reconstruction Mean Squared Error (MSE) loss, or $L_{\text{rec}}$, defined as:

\begin{equation}
    L_{\text{rec}} = \norm{x - F(x | l_{ext})}_{2}^{2},
\end{equation}

with $F(\cdot)$ corresponding to the VQ-VAE, and $l_{ext}$ corresponding to output logits of the external attribute classifier~$C_{ext}$ with regard to input $x$, that are used to condition the decoder of~$F$.

To encourage a more diverse selection of codewords, and to prevent codebook collapse (i.e., a state where only a subset of codewords are ever selected for any input), we also add a \textit{codebook diversity} loss, $L_{\text{div}}$, as proposed~by~\cite{dieleman2018challenge,baevski2020wav2vec2}:

\begin{equation}
    L_{\text{div}} = \frac{1}{GV} \sum_{g=1}^{G}\sum_{v=1}^{V} \overline{p}_{g,v} \, \text{log} \, \overline{p}_{g,v},
\end{equation}

with $V$ corresponding to the number of entries per codebook, and G corresponding to the number of codebooks in the quantization module; $\overline{p}_{g, v}$ corresponds to the per-batch average of probabilities $p_{g,v}$, defined in eq. \eqref{eq:gumbel_softmax}.

Finally, to promote target-task performance, we train the VQ-VAE for speaker identification, using a pre-trained, frozen, speaker classification layer combined with an Additive Angular Margin loss \cite{deng2019arc}, $L_{\text{aam}}$, defined as:

\begin{equation}
    L_{\text{aam}} = \frac{1}{N} \sum_{i=1}^{N} \text{log} \frac{e^{\zeta\ cos(\theta_{y_{i},i}+a)}}{\mathcal{Z}},
\end{equation}
where $\mathcal{Z}$ is defined as:
\begin{equation}
    \mathcal{Z} = e^{\zeta\ cos(\theta_{y_{i},i}+a)} + \Sigma_{j=1, j\neq i}^{c_{spk}} e^{\zeta\ cos(\theta_j, i)},
\end{equation} 

and where $N$ is the number of samples in the batch; $c_{spk}$ is the number of speaker classes; $a$ is the angular margin; $\zeta$ is a scale factor; $\theta_{y}$ is the output of the speaker classification layer for a sample~$x_i$.

The full VQ-VAE loss is then defined as:

\begin{equation}
\label{eq:total}
    L_{\text{VQ-VAE}} = \alpha L_{\text{rec}} + \beta L_{\text{div}} + \gamma L_{\text{aam}},
\end{equation}

where $\alpha$, $\beta$ and $\gamma$ are weights for each of the loss functions. This system is represented in Fig.~\ref{fig:method}, corresponding to the blue boxes. Dashed blocks correspond to components of the method that are removed at inference time.

Even though the current method, as it stands, may already have some ability to disentangle information, it does not yet explicitly promote the removal of private information. In the following sections, we detail the two approaches we use to achieve this goal: an adversarial classifier and a mutual information minimisation loss.

\subsection{Adversarial Classifier}
Following what was stated above, to promote the explicit removal of the sensitive attributes, we consider adding an adversarial classifier $C_{adv}$~\cite{goodfellow2014generative, ganin2015unsupervised}.
{The goal of this adversarial classifier is to predict the sensitive attribute from a latent representation of the VQ-VAE. 
If it can predict the attribute, then it means that the model is not removing this information. 
We want to incorporate this information when training the VQ-VAE to improve its removal ability. To this end, we train the adversarial classifier and the VQ-VAE in tandem, wherein the former will try to obtain information about the protected attribute, and the latter will try to provide as little information about it as possible.
This can be seen as a minmax game, where the VQ-VAE is trying to minimise its target loss and maximise the loss of the adversarial classifier, and the adversarial classifier is trying to minimise its own loss.

Concretely, the adversarial classifier is trained to predict the attribute from the latent representation $\mathbf{z}_q$}, whereas the VQ-VAE will be trained to prevent $C_{adv}$ from being able to correctly predict the attribute from this latent representation.
To do so, we use a gradient reversal layer (GRL)~\cite{ganin2015unsupervised}, such that $C_{adv}$ is optimised jointly with the VQ-VAE, but where the gradient corresponding to its loss is multiplied by a negative constant before being backpropagated through to the VQ-VAE.
This means that the weights of $C_{adv}$ will be adjusted to better predict the attribute, whereas the negated gradient that is passed to the VQ-VAE will adjust the weights such that it is more difficult for $C_{adv}$ to predict the attribute, and, therefore, this attribute will be hidden or absent in the latent representation of the model.

Since the attribute information will be externally fed to the decoder, adding the adversarial classifier will compel the network to learn attribute-invariant codebooks, forcing the VQ-VAE to use the external information that is fed to the decoder.

For discrete attributes, the adversarial classifier is trained using the cross-entropy loss:
\begin{equation}
    L_{adv} = - \frac{1}{c_{\text{attr}}} \sum_{i=1}^{c_{\text{attr}}} y_{{attr}_i} \, \text{log} (p_i),
\label{eq:adv_ce}
\end{equation}
where $c_{\text{attr}}$ corresponds to the number of adversarial classes, $y_{{attr}_i}$ to the attribute label, and $p_i$, the output soft-probability for class $i$ of the adversarial classifier obtained for the latent representation yielded by the quantization module, $\mathbf{z}_q$.

For continuous attributes, the MSE loss is used instead:
\begin{equation}
    L_{adv} = \norm{y_{attr} - C_{adv}(\mathbf{z}_q)}_{2}^{2}.
    \label{eq:adv_mse}
\end{equation}

The GRL, adversarial classifier and adversarial loss are represented by the dashed boxes in Fig. \ref{fig:method}.

\subsection{Mutual Information Loss}
\label{sec:mi}

%%%%%%%%%% MI %%%%%%%%%%%%
Adversarial networks have been shown to create seemingly invariant representations during adversarial training. However, these have also been shown to fail to generalise to unseen data and new classifiers trained over the new adversarial representations~\cite{elazar2018adversarial,srivastava2019privacy,noe2022bridge}.
There are several possible reasons for this to happen. For instance, during training, the adversarial classifier may no longer be able to infer the protected attribute, whereas the main network performs well for the target task. This may seem to indicate that the goal of removing the attribute was achieved. However, the adversarial network may lack the capacity (i.e., may be too simple or have too few parameters) to infer the attribute from an ``obfuscated'' latent representation, where the attribute information is hidden, thus achieving the training loss objectives without being able to actually remove information. On the other hand, one may also see adversarial training as a way of inadvertently creating \textit{adversarial examples}, i.e., data points that have suffered minute changes, but that can change a neural network's predictions~\cite{goodfellow2015explaining}.

For these reasons, in this work, we explore the usage of non-parametric nearest-neighbour-based mutual information (MI) estimators \cite{ksg,gao2018demystifying,ross2014mutual} as companion losses to the adversarial network. 
The goal of these losses is to minimise the amount of information shared between the output of the quantization module $\textbf{z}_q$ and the target attribute label $y$.
We hypothesise that, given their non-parametric nature, these losses should promote the learning of representations that are invariant to the target attribute and not simply representations that are able to "fool" the adversarial classifier.

To this end, we leverage two MI estimators: (1) the MI estimator proposed by B. Ross \cite{ross2014mutual} for mixtures of discrete and continuous random variables and (2) the 
Kraskov, St{\"{o}}gbauer and Grassberger (KSG) \cite{ksg, gao2018demystifying} estimator to estimate the MI between two continuous random variables.

The first estimator will be used as the loss between the latent representation $\textbf{z}_q$ and a discrete attribute label $y$, which, in this work, corresponds to sex information. The second estimator will be used to compute the MI loss between $\textbf{z}_q$ and $y$, in the case where $y$ is a continuous attribute, e.g. age.
In this section, we present only a high-level overview of these estimators. For further details we direct the reader to Appendix A, and to~\cite{kle,ksg, ross2014mutual, gao2018demystifying}.

\subsubsection{Mutual information estimator for discrete and continuous random variables}

The mutual information $I(Z, Y)$ between two variables $Z$ and $Y$ can be expressed in terms of the individual differential entropies and the entropy between the two random variables:
\begin{equation}
\label{eq:mi_entropy}
    I(Z, Y) = H(Z) + H(Y) - H(Z, Y).
\end{equation}

Given a set of $N$ observations taken from dataset $\mathcal{B}$ of the joint variable $M = (Z, Y)$, $m_i = (z_i, y_i)$, with $i \in 1\, ...\, N$, 
the goal of an MI estimator is to use these observations to obtain $I(Z, Y)$. 

The continuous-discrete MI estimator proposed by Ross \cite{ross2014mutual} shows that, for a discrete variable~$Y$ and a continuous variable~$Z$, the MI estimator can be obtained through a combination of nearest-neighbour entropy estimators \cite{kle}, such that:
\begin{equation}
    \hat{I}(z_i, y_i) = \psi(N) + \psi(k) - \psi(N_{y_i}) - \psi(n_{z_i}),
    \label{eq:mi_xvec}
\end{equation}

where $I(z_i, y_i)$ is the mutual information for a single observation $(z_i, y_i)$; $\psi$ corresponds to the \textit{digamma} function \cite{abramowitz1988handbook}; $k$ is a pre-specified number of neighbours; $N_{y_i}$ corresponds to number of samples in $\mathcal{B}$ with the same discrete value $y_i$; and $n_{z_i}$ is the number of samples between the continuous observation $z_i$ and its $k^{th}$ nearest-neighbour, sharing the same value $y_i$, computed using the euclidean distance.

To obtain the MI for the full set of samples, we compute the average of all $I_i(z_i, y_i)$:

\begin{equation}
    \hat{I}(X, Y) = \psi(N) + \psi(k) - \langle \psi(N_y)\rangle - \langle \psi(n_{z}) \rangle,
    \label{eq:mi_cd}
\end{equation}
where $\langle ... \rangle = \frac{1}{N}\sum_{i=1}^{N} ... $ is the average operator.

In summary, to compute the mutual information $I(z_i, y_i)$ between a vector $z_i$ and its discrete label $y_i$, we need to find $z_i$'s \textit{k\textsuperscript{th}} neighbour in a set $B$, sharing the same discrete variable. We then count the number of vectors $z$ ($n_{z_i}$) in $\mathcal{B}$, for all discrete variables $Y \neq y_i$, that are within the distance between $z_i$ and its \textit{k\textsuperscript{th}} nearest-neighbour, and the total number of observations $n_{y_i}$ with discrete value $Y = y_i$.

\begin{algorithm}[t]
\begin{algorithmic}[1]
    \STATE \textbf{Input}: batch $\mathcal{B}\!=\!(Z, Y)$ of size $N$, neighbours $k$, pairwise euclidean distance matrix (edm) function $\mbox{pdist}_{l2}(\cdot)$, $\mbox{bottom\_k}(\cdot)$ to obtain the $k^{th}$ lowest value, row-wise.
    \STATE $\mbox{edm}_Z  \gets \mbox{pdist}_{l2}(Z)$
    \STATE $N_y \gets [\,]$, $\mbox{k\_dists}\gets [\,]$
    \FOR{$y \in \{Y\}$}
       \STATE $N_y[y] \gets \#\mathcal{B}_{Z|Y=y}$
       \STATE $\mbox{k\_dists}[Y=y] \gets \mbox{bottom\_k}(\mbox{edm}_{Z|Y=y})$
    \ENDFOR
    \STATE $n_z \gets [\,]$
    \FOR{$i \in N$}
        \STATE $n_z[i] \gets 0$
        \FOR{$j \in N$}
            \STATE $n_z[i] \mathrel{+}= 1$ \textbf{if} $\mbox{edm}_z[i,j] \leq \mbox{k\_dists}[i]$
        \ENDFOR
    \ENDFOR
    \STATE $\mbox{mi} \gets \psi(N) + \psi(k) - \langle \psi(N_y) \rangle - \langle \psi(n_z) \rangle$
    \RETURN $\mbox{mi}$
\end{algorithmic}
\caption{Pseudo-code to compute \^I(Z,Y) using eq. \eqref{eq:mi_cd}}
\label{alg:mi_cd}
\end{algorithm} 

For a high-level intuition of this estimator, consider the following.
From equation \eqref{eq:mi_xvec} we can see that the MI between a vector $X$ (i.e., a speaker representation) and its discrete counterpart, $Y$ (i.e., a class label) will be lower if $n_z$ is high, and vice-versa. Note that $n_z$ is the number of samples that are not from the same class as $X$, but which are closer to $X$ than $X$ is to its $k^{th}$ nearest-neighbour belonging to the same class. Taking this into account, the MI can be seen as a measure of how well the speaker representations from each class are separated in space. If the MI is high, the vectors of each class are well separated from the other classes, and if the MI is low, then the vectors belonging to different classes will be intermixed.
Thus, using the MI as a loss will prompt the VQ-VAE to learn to create latent representations that are closer together in space independently of their attribute classes, and that do not provide discriminative information concerning their attribute classes $Y$.

This MI estimator is presented in pseudo-code in Algorithm \ref{alg:mi_cd}.

\begin{algorithm}[t]
\begin{algorithmic}[1]
    \STATE \textbf{Input}: batch $\mathcal{B}\!=\!(Z, Y)$ of size $N$, neighbours $k$, pairwise euclidean distance matrix (edm) function $\mbox{pdist}_{l2}(\cdot)$, $\mbox{bottom\_k\_idx}(\cdot)$ to obtain the row-wise index of the $k^{th}$ lowest value.
    \STATE $v_z \gets \pi^{\frac{d_Z}{2}}/\Gamma(\frac{d_Z}{2} + 1)$, $v_y \gets \pi^{\frac{d_Y}{2}}/\Gamma(\frac{d_Y}{2} + 1)$
    \STATE $\mbox{edm}_Z  \gets \mbox{pdist}_{l2}(Z)$
    \STATE $\mbox{edm}_Y  \gets \mbox{pdist}_{l2}(Y)$
    \STATE $\mbox{edm}_{ZY} \gets \mbox{pdist}_{l2}((Z, Y))$
    \STATE $\mbox{k\_dists\_idx} \gets \mbox{bottom\_k\_idx}(\mbox{edm}_{ZY})$
    \STATE $n_x \gets [\,],\, n_y \gets [\,]$
    \FOR{$i \in N$}
        \STATE $n_z[i] \gets 0,\, n_y[i] \gets 0$
        \FOR{$j \in N$}    
            \STATE $n_z[i] \mathrel{+}= 1$ \textbf{if} $\mbox{edm}_{Z}[i,j] \leq \mbox{edm}_Z[i,\! \mbox{k\_dists\_idx}[i]]$
            \STATE $n_y[i] \mathrel{+}= 1$ \textbf{if} $\mbox{edm}_{Y}[i,j] \leq \mbox{edm}_Y[i, \mbox{k\_dists\_idx}[i]]$
        \ENDFOR
    \ENDFOR
    \STATE $\mbox{mi} \gets \mbox{log}\,N + \psi(k) + \mbox{log}
    \,\frac{v_{z} v_{y}}{v_{z}+v_{y}}\!-\!\langle \mbox{log}\,n_z\!+\!\mbox{log}\,n_y \rangle$
    \RETURN $\mbox{mi}$
\end{algorithmic}
\caption{Pseudo-code to compute \^I(Z,Y) using eq. \eqref{eq:mi_cc}}
\label{alg:mi_cc}
\end{algorithm} 

\subsubsection{Mutual information estimator for continuous random variables}
For the second MI loss, between a continuous vector and a continuous attribute, we consider the use of a variant of the Kraskov, St{\"{o}}gbauer and Grassberger (KSG) MI estimator~\cite{ksg} (Algorithm 2), proposed by Gao et al.~\cite{gao2018demystifying}, where the MI is estimated through:

\begin{equation}
\begin{split}
    \hat{I}(Z, Y) & = \text{log}(N) + \psi(k) + \text{log}\frac{v_{z} v_{y}}{v_{z}+v_{y}} \\
    & - \langle \text{log}(n_z) + \text{log}(n_y) \rangle.
\end{split}
\label{eq:mi_cc}
\end{equation}

Here, $n_z$ and $n_y$ correspond to the number of points between observation $m_i = (z_i, y_i)$ and its $k^{th}$ nearest-neighbour in each marginal space ($Z$ or $Y$), being defined as the $k^{th}$ observation that is closest to the joint observation $m_i$, obtained using the euclidean distance. 
The values $v_{z}$ and $v_{y}$ correspond to the volumes of the $d_z$ and $d_y$-dimensional unit-ball, for the marginal spaces $z$ and $y$, being defined as $v = \pi^{\frac{d}{2}}/\Gamma(\frac{d}{2} + 1)$, with $\Gamma$ the \textit{gamma} function \cite{abramowitz1988handbook}.

In other words, for each pair $(z_i, y_i)$ in $\mathcal{D}$, we count the number of points ($n_z$ and $n_y$) for each random variable, that are within distances $\epsilon_{z_j}$ and $\epsilon_{y_j}$, which correspond to the distances in each marginal space between the joint observation $m_i$ and its $k_{th}$ neighbour.
As before, this MI estimator is described in pseudo-code in Algorithm \ref{alg:mi_cc}.

\subsubsection{Differentiability of the estimators}

To turn $\hat{I}(Z, Y)$ into a loss, we need to ensure that all steps in its computation are differentiable. 
Determining the $k^{th}$ closest neighbour and counting the number of data points inside a given radius are not differentiable operations. 

For simplicity, we assume that in the top-k operation (to determine the $k^{th}$ closest neighbour), gradients are only passed through to the top-k elements. In contrast, for other elements, gradients are set to zero.

On the other hand, the less or equal than comparison is implemented using a straight-through estimator of the Heaviside function:

 \begin{equation}
     (d_i \leq d_{\text{kth}}) = \text{STHeaviside}(d_{\text{kth}}-d_i).
 \end{equation}

These two adaptations allow us to use $L_{\text{MI}} = I(Z, Y)$ in combination with our model.
We positioned the loss in the same place as the adversarial classifier at the output of the quantization module. 

The MI loss is represented at the bottom of Fig.~\ref{fig:method} by a dashed circle, completing the method.

\subsection*{Full training loss}
The simplest form of our model, the VQ-VAE by itself, uses as a training loss eq. \eqref{eq:total}.

To use the adversarial classifier and loss described above, we add $L_{adv}$ to the training loss, multiplied by a weight $\delta$. Similarly, to use the MI loss (cf. eqs. \eqref{eq:mi_cd} and \eqref{eq:mi_cc}), we weight it with a constant value $\epsilon$ and add it to the remaining training losses, with the full loss becoming:

\begin{equation}
\begin{split}
    L_{\text{Total}} &= L_{\text{VQ-VAE}} + + \delta L_{adv} + \epsilon L_{\text{MI}} \\
    &= \alpha L_{\text{rec}} + \beta L_{\text{div}} + \gamma L_{\text{aam}} + \delta L_{adv} + \epsilon L_{\text{MI}}.
\end{split}
    \label{eq:full_loss}
\end{equation}

\section{Experimental Setup}
\label{sec:exp}

%%%%%%%%%% Experimental Setup %%%%%%%%%%%%

\subsection{Experiments}

As mentioned in Section \ref{sec:intro}, two speaker attributes are considered, sex and age, which should be removed from speaker representations using the method described in the previous section. This is done with two different models, one for each attribute, each trained using the losses that are appropriate to discrete (i.e., sex) or continuous labels (i.e., age).

For the proposed method to be validated, it is necessary to show that it fulfils the objectives detailed at the beginning of Section \ref{sec:method} section: a) the method should be able to remove and manipulate attribute information, and b) the method should have little impact on the target task (speaker verification).

To validate both of these conditions, we conduct an extensive set of experiments:

\begin{table*}[ht]
\centering
\captionsetup{justification=centering}
\caption{Data partitions for the VoxCeleb and LibriTTS datasets.}
    \begin{tabular}{cc|rrr|rrr}
    \toprule
     \multirow{2}{*}{Source dataset} & \multirow{2}{*}{Partition} & \multicolumn{3}{c|}{$\#$Speakers} & \multicolumn{3}{c}{$\#$Utterances} \\
     
     & & \multicolumn{1}{c}{Male} & \multicolumn{1}{c}{Female} & \multicolumn{1}{c|}{Total} & \multicolumn{1}{c}{Male} & \multicolumn{1}{c}{Female} & \multicolumn{1}{c}{Total} \\ \midrule

    \multirow{4}{*}{VoxCeleb} & train\_vox\_spk & 4,347 & 2,858 & 7,205 & 1,459,045 & 887,649 & 2,346,694 \\ \cmidrule{2-8}
    & train\_vox\_vq  & 2,572 & 2,572 & 5,144 & 467,870 & 412,225 & 880,095 \\
    & train\_vox\_att & 209 & 191  & 400 & 37,444  & 29,835 & 67,279 \\ 
    & test\_vox\_att  & 91 & 46 & 137 & 24,598 & 9,511 & 34,109 \\ \midrule

    \multirow{3}{*}{LibriTTS} & train\_libri\_vq  & 600 & 560 & 1,160 & 100,364 & 104,680 & 205,044 \\
    & train\_libri\_att & 474 & 430 & 904 & 55,619 & 60,881 & 116,500 \\
    & test\_libri\_att  & 164 & 162 & 326 & 20,274 & 23,536 &  43,810 \\ \midrule

    Vox+Libri & train\_vox\_libri\_vq & 3173 & 3132 & 6305 & 209,286 & 200,623 & 409,909 \\
    
     \bottomrule
    \end{tabular}
    \label{tab:data}
\end{table*}

\begin{enumerate}
    \item An ablation study is conducted to compare the performance of a simple VQ-VAE with versions of the same VQ-VAE to which the adversarial loss $L_{adv}$ or the mutual information loss, $L_{MI}$, were added, and finally, when both losses are used in combination.
    This study concerns both the sex and age attributes, and we report results in terms of privacy (i.e., the ability to remove the attribute) and utility (i.e., speaker verification performance).

    \item The results that were obtained for the sex attribute are compared to the method of No\'e  et al. \cite{noe2022bridge}, the Normalising Flow zero Log-Likelihood Ratio (NFzLLR). This method was selected because it is a good representative of the state-of-the-art for attribute removal from speaker representations and because it is the work that has the closest evaluation methodology to our own.

    \item We perform cross-domain experiments to understand how robust the proposed method is to domain changes. To do so, we use an out-of-domain dataset with which we replace (1) the test data, (2) the training data of the attribute classifier and (3) the training data of the VQ-VAE itself. 

    \item We test the manipulation capabilities of our method for both attributes. To this end, we treat the externally provided attribute information as the true labels and measure the performance of pre-trained (i.e., trained on unprotected data) sex and age classifiers in classifying the false information. This way, we are able to obtain an indication of whether the proposed method did indeed replace the true attribute with the fake one.

    \item Due to a lack of age-labelled speech data sources, the cross-domain experiments are only applied to the sex information removal models.

\end{enumerate}

All experiments are reported for both \textit{ignorant} and \textit{informed} attackers, except for the attribute manipulation experiment, where we only consider the \textit{ignorant} scenario. 

\subsection{Data}
\label{sec:data}
Four datasets are used in our experiments: VoxCeleb~\cite{nagrani2020voxceleb}; LibriTTS~\cite{zen2019libritts}; an age annotated partition of VoxCeleb named AgeVoxCeleb~\cite{tawara2021age}; and a Portuguese version of the VoxCeleb corpus, VoxCelebPT~\cite{mendonca2022voxcelebpt}, which contains annotations on both the speakers' sex and ages.
Next, we describe each of these datasets, as well as why and how they are used for the experiments described above.

\subsubsection{VoxCeleb}

VoxCeleb \cite{nagrani2020voxceleb} is the primary source of data for the experiments presented in this work.
This corpus includes recordings of 7,363 speakers of multiple ethnicities, accents, occupations, age groups and languages, having English as the most prevalent language. It is composed of short clips taken from interviews uploaded to YouTube. The corpus is composed of two parts, \textit{VoxCeleb 1 and 2}, both subdivided into \textit{dev} and~\textit{test}.
This corpus is one of the most widely used publicly available corpora for speaker recognition tasks. It is also one of the largest corpora for this task, both in terms of the number of speakers and individual utterances per speaker, presenting a large variety of, often noisy, recording conditions. Moreover, its test set is often used as a benchmark to evaluate new speaker recognition models. These characteristics, as well as the fact that many pre-trained speaker embedding extraction models are trained on this dataset, make it ideal for the experiments performed in this work.

We use four data partitions, described in detail in Table \ref{tab:data}, three of which are used for training the different components of our method, and the fourth is used for testing.

The first partition -- \textit{train\_vox\_spk} -- corresponds to the data used to train the speaker embedding extraction model and corresponds to the full \textit{dev} set of VoxCeleb (1+2), with 7,205 speakers.

The second partition -- \textit{train\_vox\_vq} -- is used to train the VQ-VAE for the sex attribute. It uses a subset of 5,144 speakers (balanced by sex), taken from the \textit{dev} set of VoxCeleb (1+2). 
This partition is also used to train the external sex classifier, from which we extract the logits used to condition the VQ-VAE's decoder.

The third partition -- \textit{train\_vox\_att} -- is composed of a second set of 400 speakers, also taken from the \textit{dev} set of VoxCeleb, having no speaker overlap with the partition used to train the VQ-VAE. This partition is used to train the sex classifiers that evaluate the privacy capabilities of our method.

All sex attribute-related experiments are evaluated using a combination of the \textit{test} sets of VoxCeleb 1 and 2 -- \textit{test\_vox\_att}. 
However, Nagrani et al.~\cite{nagrani2020voxceleb} warn that there may be a speaker overlap between the VoxCeleb 1 \textit{dev} and \textit{test} partitions with VoxCeleb 2 \textit{test}. We manually checked the speakers in VoxCeleb 2 \textit{test} and found 21 speakers that were present in VoxCeleb 1. These speakers were removed from the test set to avoid contamination from the training data. This resulted in a final set of 137 test speakers.

Speaker verification performance is evaluated using VoxCeleb 1's original trial pairs, taken from VoxCeleb 1's test partition, corresponding to a set of 40 speakers, 4,874 utterances and a total of 37,720 trials.

\subsubsection{LibriTTS}

\begin{table}[t]
\centering
\caption{In-domain and cross-domain experiments.}
\begin{tabular}{@{}c|c|c|c@{}}
\toprule
\textbf{Partition} & Train VQ-VAE & Train $C_{att}$ & Test  $C_{att}$ \\ \midrule
\multirow{8}{*}{\textbf{Domain}} & \multirow{4}{*}{VoxCeleb} & \multirow{2}{*}{VoxCeleb} & VoxCeleb \\ \cmidrule(l){4-4} 
 & & & LibriTTS \\ \cmidrule(l){3-4} 
 & & \multirow{2}{*}{LibriTTS} & VoxCeleb \\ \cmidrule(l){4-4} 
 & & & LibriTTS \\ \cmidrule(l){2-4} 
 & \multirow{4}{*}{LibriTTS} & \multirow{2}{*}{VoxCeleb} & VoxCeleb \\ \cmidrule(l){4-4} 
 & & & LibriTTS \\ \cmidrule(l){3-4} 
 & &\multirow{2}{*}{LibriTTS} & VoxCeleb \\ \cmidrule(l){4-4}
 & & & LibriTTS \\ \bottomrule
\end{tabular}
\label{tab:crossdomain}
\end{table}

\begin{table*}[t]
\centering
\captionsetup{justification=raggedright}
\caption{Data partitions for AgeVoxCeleb and VoxCelebPT.}

    \begin{tabular}{c|c|c|rrrrrr|r}
    \toprule     
    \multicolumn{1}{c|}{Source dataset} & Partition & Utt./Spk. & $<=$20 & 30-39 & 40-49 & 50-59 & 60-69 & $>=$70 & Total \\ \midrule

    \multirow{2}{*}{AgeVoxCeleb} 
    & \multirow{2}{*}{train\_agevox} 
    & \#Speakers      & 1,531 & 1,773 & 1,292 & 921 & 567 & 217 & 4,220 \\
    & & \#Utterances  & 26,970 & 34,856 & 30,548 & 25,751 & 17,686 & 5,757 & 141,568 \\ \midrule

    \multirow{2}{*}{VoxCelebPT} 
    & \multirow{2}{*}{test\_voxpt} & \#Speakers & 7 & 12 & 14 & 7 & 6 & 5 & 51 \\
    & & \#Utterances  & 3,855 & 6,610 & 7,722 & 3,402 & 3,034 & 2,113 & 26,736 \\
    
     \bottomrule
    \end{tabular}
    \label{tab:data_age}
\end{table*}

Our second main source of data is LibriTTS \cite{zen2019libritts}. This dataset is an adaptation of the LibriSpeech corpus -- a corpus of read speech, fully in English, taken from audiobooks -- wherein the data was processed to be suitable for text-to-speech tasks. The complete LibriTTS corpus amounts to a total of 586.5 hours, containing 2,456 speakers.

In our cross-domain study for the sex attribute, we use this dataset to assess how well our model generalises to unseen domains. LibriTTS is comprised of read speech, recorded under controlled conditions, which makes it starkly different from VoxCeleb, where the speech recordings are noisy and contain spontaneous speech, making this dataset an ideal source of out-of-domain data.
The motivation for this experiment comes partly from the fact that the VQ-VAE, the sex attribute classifier, and the speaker embedding extraction model are all trained on VoxCeleb, possibly giving us biased results.

For the above reasons, in order to assess the impact of domain changes, we perform a total of 8 experiments using different combinations of VoxCeleb and LibriTTS. These include replacing the data used to train the VQ-VAE, the data used to train the attribute classifier and the test data. These experiments, and the in-domain experiments, are summarised in Table \ref{tab:crossdomain}, where each line corresponds to one experiment, and each column corresponds to the different tasks for which the data is used. 

To perform these experiments, we use three LibriTTS partitions: \textit{train\_libri\_vq}, \textit{train\_libri\_att} and \textit{test\_libri\_att}. The first is used to train the VQ-VAE, the second is used to train attribute classifiers, and the third is used as a test set. The \textit{train\_libri\_vq} partition comprises data taken from LibriTTS' train-other-500 partition; \textit{train\_libri\_att} uses data taken from train-clean-360 and, \textit{test\_libri\_att} combines data taken from train-clean-100, dev-clean and test-clean. Each speaker is present only in a single partition.

Finally, we use \textit{train\_vox\_libri\_vq} to train the VQ-VAE, in one of the cross-domain scenarios, where 50\% of the VQ-VAE's training partition is composed of data taken from LibriTTS, and 50\% is taken from VoxCeleb. Specifically, the subset of LibriTTS data corresponds to \textit{train\_libri\_vq}, and the subset of VoxCeleb corresponds to \textit{train\_vox\_vq}, with the number of samples downsampled to match the size of \textit{train\_libri\_vq}.
More details for each partition can be found in Table~\ref{tab:data}.

\subsubsection{AgeVoxceleb \& VoxCelebPT}
For our age-related experiments, we use two datasets: AgeVoxCeleb~\cite{tawara2021age} and VoxCelebPT~\cite{mendonca2022voxcelebpt}. The full details of the partitions used in our experiments can be found in Table~\ref{tab:data_age}.

AgeVoxCeleb is a subset of VoxCeleb 2 that has been annotated with speaker age labels, obtained by cross-checking birth years found online, with video recording and broadcasting dates. This dataset is composed of 4,976 speakers and 21,707 utterances, with several speakers having multiple utterances at different ages. 
It is, to the best of our knowledge, the largest publicly available age-labelled speech corpus. This, and the fact that it is a subset of VoxCeleb 2, prompted us to select this dataset for our age-related experiments.

VoxCelebPT~\cite{mendonca2022voxcelebpt} is a Portuguese version of VoxCeleb, containing recordings of 51 Portuguese celebrities obtained online. This corpus amounts to a total of 26,736 utterances, manually annotated with sex and age labels. In this work, we use a subset of this corpus, containing 25,929 utterances with a minimum length of 1s.

In our experiments, we used AgeVoxCeleb -- train\_agevox -- as the training data for the VQ-VAE and the age classifier. 
Given the small size of this dataset, when compared to the one used for sex classification, we decided to use the same partition for both the VQ-VAE and the attribute classifier, as our preliminary experiments with smaller partitions showed poor performance for age regression.
VoxCelebPT is used as held-out test data -- test\_voxpt. 
Even though it is also comprised of interviews, under a wide variety of recording conditions -- the reason for which it was selected -- this dataset can also be considered out-of-domain data since it only contains recordings of European Portuguese.

\subsection{Evaluation}
\label{sec:eval}
To evaluate the performance of our method in terms of privacy concerning sex information, we use two binary classification metrics:
Unweighted Average Recall (UAR) and Area Under the Precision Recall Curve~(AUPRC). The UAR reflects the performance of a classifier on a fixed threshold, whereas the AUPRC reports the average classifier performance over all possible classification thresholds.
Both have a chance level of 50\% for binary classification with imbalanced datasets. These metrics should be as close to 50\% as possible for privatised speaker embeddings and as close to 100\% as possible for the original, non-protected vectors.

For comparison with the work of~\cite{noe2022bridge}, we also report two Privacy Zebra metrics \cite{nautsch2020privacyzebra}. The first Zebra metric is
D$_\text{ECE}$, the \textit{expected privacy disclosure} which compares the amount of information provided by the oracle-calibrated output log-probabilities of a classifier and that of a non-informative posterior.
The second Zebra metric we consider is the llr$_{\text{max}}$, which measures the worst-case privacy disclosure among the test data by selecting the highest log-likelihood ratio for a single sample over oracle calibrated log-probabilities.
For both metrics, values close to zero correspond to better privacy protection.

For age, we use the Concordance Correlation Coefficient (CCC) and Pearson's Correlation Coefficient (PCC) as metrics. The CCC measures whether the classifier's output exactly matches the provided labels, being a conservative estimate of the classifier's performance.
On the other hand, the PCC measures correlation up to a linear transformation, corresponding to a more optimistic view of the classifier's performance.

Speaker verification performance is evaluated in terms of Equal Error Rate~(EER) and of the minimum of the Detection Cost Function~(minDCF). We use the cosine similarity between two embeddings as the scoring method.

\subsection{Implementation details}

We use SpeechBrain's pre-trained ECAPA-TDNN \cite{desplanques2020ecapa, speechbrain} as our speaker embedding extractor. This model was trained on the development set VoxCeleb 1+2, as described in Section \ref{sec:data}. Speaker embeddings extracted from the ECAPA-TDNN have size 192. The complete architecture of this network can be found in \cite{desplanques2020ecapa}.

The encoder and decoder modules of the VQ-VAE (for both attributes) are composed of 3 hidden layers, all of size 512, except for the $3^{\text{rd}}$ layer of the encoder, which has size $h$ = $128$, to create a bottleneck. 
The decoder has an output layer of size $n$ = $192$ to match the input embeddings.
The quantization module is composed of $G$ = $64$ codebooks, with $V$ = $128$ entries of size $(e/G)$ = $4$. 
The quantization module linear transformation layer has dimension $q$ = $256$, whereas the external logits linear layer has size $w$ = $4$ to match the size of the codewords.
In total, our model amounts to $\sim1$M parameters.

Attribute classifiers are composed of 2 hidden layers of size 128 and an output layer of size $c_{\text{attr}}$, corresponding to the number of classes of the attribute at hand -- 2 for sex and 1 for age.
The adversarial classifier is composed of an input Batch Normalisation (BN) layer \cite{ioffe2015batch}, 3 hidden layers of size 128, and an output layer of size $c_{\text{attr}}$.
All hidden layers consist of a linear layer, a Leaky-ReLU activation, and a BN layer.

Speaker classification, to compute the $L_{aam}$ loss, is performed with a linear layer, pre-trained with the same data used to train the VQ-VAE. This layer is frozen to force the model to ensure perfect reconstruction.

All models were trained with Adam \cite{kingma2014adam}, using a one-cycle learning rate (\textit{lr}) policy \cite{smith2019super}.
VQ-VAE models were trained for 100 epochs, using a start \textit{lr} of 8$\times$$10^{-4}$, and a maximum of 0.01, dropout probability of 0.1 and a batch size of 128; attribute classifiers were trained for 20 epochs, with a start \textit{lr} of $10^{-5}$, and a maximum of 5$\times$$10^{-5}$, a dropout probability of 0.3 and a batch size of 64.
When training the VQ-VAE for the sex attribute, we ensure batches are always balanced in terms of sex, per sample.

For all experiments, except for the manipulation experiment, when testing the VQ-VAE, the decoder is fed with the same \textit{fake} attribute. This fake attribute corresponds to the mean value of the logits outputted by the pre-trained external attribute classifier, computed over the full training set.
The reasoning behind this selection is that, by providing the mean logits for the attribute, we are providing a possible attacker with the least possible amount of information \cite{noe2021adversarial}.

When performing the attribute manipulation experiment, the VQ-VAE is fed random attribute logits that follow a simple Gaussian distribution to ensure they fall within the observed range of logit values. We select random attribute logits in this experiment to ensure that there is sufficient coverage of possible attribute values when testing the performance of the pre-trained classifier over these \textit{fake} attributes.

Both MI losses use $k=4$ neighbours and the $l^2$-norm as the distance metric. $L_{aam}$ has a margin of $m=0.2$ and a scale factor of $s=30$.

For all VQ-VAE models, the reconstruction loss $L_{rec}$ has weight $\alpha = 1.0$, the codebook diversity loss $L_{div}$ has weight $\beta = 0.1$, and the Additive Angular Margin loss $L_{aam}$ has weight $\gamma = 1.0$.

For the sex attribute, the VQ-VAE is trained with $\delta = 1000$ when using only the adversarial classifier, with $\epsilon = 100$ when using only the MI loss, and $\delta = \epsilon = 10$ when both losses are used.
For the age attribute, the VQ-VAE is trained with $\delta = 1$ when using only the adversarial classifier, with $\epsilon = 100$ when using only the MI loss, and $\delta = 1$, $\epsilon = 10$ when the two losses are used in combination.
This selection was made through a hyper-parameter search, using powers of ten in the range of $[0.1, 1000]$ as the weights for each loss.

\begin{table*}[ht]
\centering
\captionsetup{justification=raggedright}
\caption{Results regarding the removal of sex information for ignorant attackers.}
\begin{tabular}{l|cc|cc|cc}
\toprule
\multicolumn{1}{c}{\multirow{2}{*}{\textbf{Model}}} & \multicolumn{2}{c}{\textbf{Speaker Verification Metrics}}            & \multicolumn{2}{c}{\textbf{Sex Classification Metrics}}     & \multicolumn{2}{c}{\textbf{Sex Privacy Metrics}}      \\ \cmidrule(l){2-7} 
\multicolumn{1}{c}{} 
& \textbf{EER (\%) $\downarrow$} 
& \multicolumn{1}{c|}{\textbf{minDCF $\downarrow$}} 
& \textbf{AUPRC (\%) $\downarrow$} 
& \multicolumn{1}{c|}{\textbf{UAR (\%)} $\downarrow$}
& $\text{\textbf{D}}_\text{\textbf{ECE}}$$\downarrow$      
& $\text{\textbf{llr}}_\text{\textbf{max}}$$\downarrow$    \\ \midrule

Original data       & 0.88          & 0.0011 & 99.40 $\pm$ 0.11 & 97.74 $\pm$ 0.28 & 0.649 $\pm$ 0.007 & 3.444 $\pm$ 0.176 \\ \midrule
NFzLLR \cite{noe2022bridge} & 4.89  & 0.0043 & \textbf{51.29 $\pm$ 0.96} & 51.72 $\pm$ 0.66 & \textbf{0.002 $\pm$ 0.001} & \textbf{0.633 $\pm$ 0.245} \\ \midrule
VQ-VAE              & \textbf{1.44} & 0.0021 & 82.35 $\pm$ 1.09 & 73.82 $\pm$ 1.35 & 0.218 $\pm$ 0.014 & 2.262 $\pm$ 0.227 \\
VQ-VAE + MI         & 2.12          & 0.0026 & 60.54 $\pm$ 1.30 & 56.11 $\pm$ 1.31 & 0.039 $\pm$ 0.009 & 1.690 $\pm$ 0.394 \\
VQ-VAE + ADV        & 2.45          & 0.0029 & 56.72 $\pm$ 0.84 & 54.76 $\pm$ 0.78 & 0.016 $\pm$ 0.004 & 0.883 $\pm$ 0.327 \\ \midrule
VQ-VAE + ADV + MI   & 1.48          & \textbf{0.0019} & 52.92 $\pm$ 0.92 & \textbf{50.91 $\pm$ 0.60} & 0.005 $\pm$ 0.002 & 0.761 $\pm$ 0.289 \\ \bottomrule
\end{tabular}
\label{tab:results_bat}
\end{table*}
\begin{table*}[ht]
\centering
\captionsetup{justification=raggedright}
\caption{Results regarding the removal of sex information for informed attackers.}
\begin{tabular}{l|cc|cc}
\toprule
\multicolumn{1}{c}{\multirow{2}{*}{\textbf{Model}}} & \multicolumn{2}{c|}{\textbf{Sex Classification Metrics}} & \multicolumn{2}{c}{\textbf{Sex Privacy Metrics}} \\ \cmidrule{2-5} 
\multicolumn{1}{c}{} 
& \textbf{AUPRC} (\%) $\downarrow$                
& \textbf{UAR} (\%) $\downarrow$               
& $\text{\textbf{D}}_{\text{\textbf{ECE}}}$ $\downarrow$             
& $\text{{\textbf{llr}}}_\text{\textbf{max}}$ $\downarrow$       \\ \midrule
Original data     & 99.40 $\pm$ 0.11 & 97.74 $\pm$ 0.28 & 0.649 $\pm$ 0.007 & 3.444 $\pm$ 0.176  \\ \midrule
NFzLLR \cite{noe2022bridge} 
                  & 74.59 $\pm$ 0.85 & 71.36 $\pm$ 0.68 & 0.138 $\pm$ 0.008 & 1.839 $\pm$ 0.177 \\ \midrule
VQ-VAE            & 90.89 $\pm$ 0.68 & 85.67 $\pm$ 0.70 & 0.367 $\pm$ 0.013 & 2.844 $\pm$ 0.158 \\
VQ-VAE + MI       & 72.78 $\pm$ 1.09 & 70.31 $\pm$ 0.89 & 0.132 $\pm$ 0.010 & 2.345 $\pm$ 0.197 \\
VQ-VAE + ADV      & 63.18 $\pm$ 0.84 & 62.62 $\pm$ 0.69 & 0.052 $\pm$ 0.005 & 1.474 $\pm$ 0.195 \\ \midrule
VQ-VAE + ADV + MI & \textbf{57.41 $\pm$ 0.67} & \textbf{57.71 $\pm$ 0.87} & \textbf{0.021 $\pm$ 0.004} & \textbf{1.145 $\pm$ 0.255} \\ \bottomrule
\end{tabular}
\label{tab:results_aat}
\end{table*}

To train the NFzLLR model, we use the authors' original implementation \cite{noe2022bridge}, available online\footnote{\url{https://github.com/LIAvignon/bridge-features-evidence}}. We use the same data partitions that we use to train and test our models. Since a hyper-parameter search for this model was out of the scope of this work, we tried the two hyper-parameter configurations used by the authors in \cite{noe2022bridge, noe2023hiding}. By comparing the results for both configurations, we determined that the hyper-parameters used in \cite{noe2023hiding} provided the best results in terms of privacy. Moreover, these hyper-parameters were selected for ECAPA-TDNN speaker embeddings, the same as the one used in this work. Nonetheless, in our experiments, the hyper-parameter configuration of \cite{noe2022bridge} provided better results in terms of speaker verification.

All attribute classification (or regression) results were obtained by training the attribute classifiers 25 times, with different random initialisations. All privacy metrics are reported as the mean $\pm$ standard deviation, computed over all runs. Speaker verification results are obtained over a single run, as there is no source of randomness in this experiment~\footnote{The code required to reproduce the experiments presented in this paper can be found in \url{https://github.com/fsepteixeira/Filter-VQVAE}.}.

\section{Results}
\label{sec:results}

This section provides the results of our experiments.
In the first two subsections, we report results for the sex and age removal experiments (experiments 1 and 2).
After, we report the results of the experiments regarding the manipulation of sex information (experiment 3) and the cross-domain experiments (experiment 4).

\subsection{Removal of sex information}

The results for the removal of sex information can be found in Table \ref{tab:results_bat} for the ignorant attacker and in Table \ref{tab:results_aat} for the informed attacker. In both tables, down-pointing arrows mean that lower values are better.

In each table, we report sex classification results for the \textit{Original} (i.e., non-transformed) speaker embeddings, as well as the results obtained for \textit{NFzLLR}~\cite{noe2022bridge}.
This is followed by the results of the ablation study, where we include results for the VQ-VAE trained without any adversarial loss, for the combination of the VQ-VAE with either the MI or the adversarial loss, and for the complete method, using a combination of both losses.

From Tables \ref{tab:results_bat} and \ref{tab:results_aat}, we can observe that each component of our method provides consistent improvements over the simple VQ-VAE.
By adding the MI loss to the method, we observe a sex classification performance degradation of more than 15\% for UAR and AUPRC when compared to the VQ-VAE for both attacker settings.
When adding the adversarial classifier and loss, we see a similar improvement to that of the MI loss for the \textit{ignorant attacker} setting. However, for the \textit{informed attacker}, the degradation is much more pronounced, over 20\% UAR and AUPRC, showing that the adversarial classifier provides a better ability to remove sex information. This is to be expected, as the adversarial loss is parametric -- it is based on a classifier -- whereas the MI loss is non-parametric.

Notably, the results show that combining the adversarial classifier with the MI loss also yields the best overall performance in terms of privacy protection. This proves that these two approaches complement each other in terms of information removal, validating our method. 
In terms of the Zebra metrics, the results follow a similar trend, with each component providing consistent improvements over the baseline.

One should also note that none of the considered methods is able to remove sex information entirely. This can be seen in the results for the \textit{informed attacker}, where the sex classification performance reaches values close to 60\% UAR and AUPRC.

\begin{table*}[ht]
\centering
\captionsetup{justification=raggedright}
\caption{Results for age regression for both ignorant and informed attackers.}
\begin{tabular}{l|cc|cc|cc}
\toprule
\multicolumn{1}{c}{\multirow{3}{*}{\textbf{Model}}} 
& \multicolumn{2}{c}{\multirow{2}{*}{\textbf{Speaker Verification Metrics}}} & \multicolumn{4}{c}{\textbf{Age Regression Metrics}}  \\ \cmidrule(l){4-7} 

\multicolumn{1}{c}{} 
& \multicolumn{2}{c}{} & \multicolumn{2}{c|}{\textbf{Ignorant Attacker}} & \multicolumn{2}{c}{\textbf{Informed Attacker}} \\ \cmidrule(l){2-7} 

\multicolumn{1}{c}{} 
& \textbf{EER (\%) $\downarrow$}  & \multicolumn{1}{c|}{\textbf{minDCF} $\downarrow$} & \textbf{CCC} $\downarrow$ & \textbf{PCC} $\downarrow$  & \textbf{CCC} $\downarrow$ & \textbf{PCC} $\downarrow$ \\ \midrule
Original data & 0.88 & 0.0011 & 0.681 $\pm$ 0.005 & 0.753 $\pm$ 0.003 & 0.681 $\pm$ 0.005 & 0.753 $\pm$ 0.003 \\ \midrule
VQ-VAE        & \textbf{1.74} & \textbf{0.0018} & 0.194 $\pm$ 0.009 & 0.370 $\pm$ 0.015 & 0.198 $\pm$ 0.013 & 0.315 $\pm$ 0.021 \\
VQ-VAE + MI   & 1.97 & 0.0024 & 0.147 $\pm$ 0.011 & 0.279 $\pm$ 0.020 & 0.160 $\pm$ 0.012 & 0.259 $\pm$ 0.018 \\
VQ-VAE + ADV  & 2.68 & 0.0027 & 0.117 $\pm$ 0.010 & 0.229 $\pm$ 0.020 & 0.119 $\pm$ 0.011 & 0.184 $\pm$ 0.017 \\ \midrule
VQ-VAE + ADV + MI & 4.24 & 0.0039 & \textbf{0.042 $\pm$ 0.009} & \textbf{0.084 $\pm$ 0.018} & \textbf{0.101 $\pm$ 0.012}  & \textbf{0.165 $\pm$ 0.020}  \\ \bottomrule
\end{tabular}
\label{tab:results_age}
\end{table*}

For the target task, speaker verification, the results show that the proposed method introduces an absolute degradation of 1.2\% and 1.6\% EER for the VQ-VAE trained with the MI loss and ADV loss, respectively, when compared to the original vectors. On the other hand, the combination of the two losses introduces a degradation of only 0.6\% EER. A possible reason for this is the fact that, for this model, the weights of both losses are set to 10.0, whereas for the MI or ADV-only models, the corresponding weights are 100.0 and 1000.0. For this reason, these losses will have a much higher impact than the MSE and $L_{aam}$ losses, where the weights are set to 1.0 and 0.1. This set of weights was selected because it provided the best performance in terms of privacy.

When comparing our approach to that of \cite{noe2022bridge}, we see that our complete method (VQ-VAE+ADV+MI) is on par with the NFzLLR for privacy protection for the ignorant attacker, in terms of the classification metrics, whereas for the Zebra metrics, our method provides worse privacy results.
This may be because the NFzLLR model was specifically developed to minimise the amount of information disclosed to an attacker -- the log-likelihood ratio between the two classes is set precisely to 0 -- which is exactly what is measured by the Zebra metrics. 
In our model, we are providing the mean "attribute" for all samples, which does not necessarily carry zero information about any class, i.e., pre-trained classifiers may interpret the mean as one class instead of no class.

Contrarily, considering the informed attacker, our method shows a much better ability to protect sex information, with a difference of more than 10\% for the classification metrics. For the Zebra metrics, our method also shows a marked improvement over the NFzLLR.
In addition, the NFzLLR shows a much higher degradation for speaker verification, being close to 5\% EER, as opposed to our 1.5\%. 

However, these results differ from those provided in \cite{noe2022bridge}, where the model had much better behaviour against informed attackers and where the degradation introduced by the model was much lower. One possible explanation for the privacy results may be the fact that in \cite{noe2022bridge}, only 71 speakers and 17,735 utterances were used to train the attribute classifier, whereas, in this work, we use 400 speakers and 67,279 utterances. For the results in terms of speaker verification, a possible reason may be the fact that, unlike \cite{noe2022bridge}, we use cosine scoring instead of Probabilistic Linear Discriminant Analysis (PLDA) scoring to perform speaker verification. 
Nevertheless, it is necessary to state that no hyper-parameter tuning was made for the NFzLLR and that better results could potentially be obtained by performing a hyper-parameter search.

\subsection{Removal of age information}

The results concerning the removal of age information can be found in Table \ref{tab:results_age}.
Similar to the sex attribute experiment, we observe a consistent improvement with each loss being added to the model, with the combination of the MI and adversarial losses providing the best results in both attacker settings.

In particular, we observe a 90\% relative improvement in terms of privacy for both correlation metrics in the ignorant attacker, a value that is reduced to between 80-85\% for the informed attacker.
When compared to the results for sex, this improvement is much higher. For the sex attribute, the relative improvement was close to 40\% AUPRC and UAR for the ignorant attacker and close to 45\% for the informed attacker. 
This shows that our method is able to generalise to continuous attributes successfully.

Nevertheless, for this attribute, the informed attacker does not provide a performance improvement over the ignorant attacker, as was observed for the sex information, for the cases where the VQ-VAE is only combined with one of the two losses.
Moreover, we must also note that for the best privacy model, the ASV performance suffers from a degradation of 3.4\% EER, which is much larger than for the sex attribute, where the degradation was kept at 0.6\%.

A possible reason for these two phenomena may be the amount of data used to train the VQ-VAE in this experiment, which corresponds to about one-eighth of the amount of data used for the sex attribute experiment. The degradation of the speaker representations that is indicated by the poor ASV performance may also affect the age regression model, such that even when it is trained over the transformed representations, it is not able to generalise properly to unseen data.
As such, we hypothesise that observing such a lower amount of data during training may have prevented the model from achieving a better trade-off between privacy and utility, with the model degrading the signal more in favour of privacy.

\subsection{Attribute manipulation results}

\begin{table*}[ht]
\centering
\captionsetup{justification=raggedright}
\caption{Results for the proposed methods for sex information manipulation within the speaker representations.}
\begin{tabular}{l|cc|cc}
\toprule
\multicolumn{1}{c|}{\multirow{2}{*}{\textbf{Model}}} & \multicolumn{2}{c|}{\textbf{Speaker Verification Metrics}} & \multicolumn{2}{c}{\textbf{Sex Classification Metrics}} \\ \cmidrule(l){2-5} 
\multicolumn{1}{c|}{} & \textbf{EER (\%) $\downarrow$} & \textbf{minCLLR} $\downarrow$ & \textbf{AUPRC (\%) $\uparrow$} & \textbf{UAR (\%) $\uparrow$}  \\ \midrule
Original data     & 0.88            & 0.0011              & 99.40 $\pm$ 0.11          & 97.74 $\pm$ 0.28 \\ \midrule
VQ-VAE            & 1.13 $\pm$ 0.04 & 0.0016 $\pm$ 0.0001 & 91.94 $\pm$ 0.34          & 85.09 $\pm$ 0.85 \\
VQ-VAE + MI       & 1.24 $\pm$ 0.05 & 0.0016 $\pm$ 0.0001 & 95.13 $\pm$ 0.74          & 86.98 $\pm$ 0.84 \\
VQ-VAE + ADV      & 1.65 $\pm$ 0.05 & 0.0022 $\pm$ 0.0002 & 96.94 $\pm$ 0.15          & \textbf{90.97 $\pm$ 0.83} \\ \midrule
VQ-VAE + ADV + MI & \textbf{1.03 $\pm$ 0.04} & \textbf{0.0014 $\pm$ 0.0001} & \textbf{97.23 $\pm$ 0.18} & 90.23 $\pm$ 0.68          \\ \bottomrule
\end{tabular}
\label{tab:results_manipulation}
\end{table*}

\begin{table*}[ht]
\centering
\captionsetup{justification=raggedright}
\caption{Results for the proposed methods for age information manipulation within the speaker representations.}
\begin{tabular}{l|cc|cc}
\toprule
\multicolumn{1}{c|}{\multirow{2}{*}{\textbf{Model}}} & \multicolumn{2}{c|}{\textbf{Speaker Verification Metrics}} & \multicolumn{2}{c}{\textbf{Age Regression Metrics}} \\ \cmidrule(l){2-5} 
\multicolumn{1}{c|}{} & \textbf{EER (\%) $\downarrow$} & \textbf{minCLLR} $\downarrow$ & \textbf{CCC $\uparrow$} & \textbf{PCC $\uparrow$}  \\ \midrule
Original data     & 0.88  & 0.0011 & 0.681 $\pm$ 0.005 &  0.753 $\pm$ 0.003  \\ \midrule
VQ-VAE            & \textbf{1.56} $\pm$ \textbf{0.03} & 0.0015 $\pm$ 0.0001 & 0.883 $\pm$ 0.007 & 0.889 $\pm$ 0.003 \\
VQ-VAE + MI       & 1.72 $\pm$ 0.02 & 0.0021 $\pm$ 0.0001 & 0.898 $\pm$ 0.008 & 0.908 $\pm$ 0.003 \\
VQ-VAE + ADV      & 2.41 $\pm$ 0.04 & 0.0024 $\pm$ 0.0001 & \textbf{0.915} $\pm$ \textbf{0.007} & 0.926 $\pm$ 0.002 \\ \midrule
VQ-VAE + ADV + MI & 3.71 $\pm$ 0.04 & 0.0034 $\pm$ 0.0001 & 0.914 $\pm$ 0.014 & \textbf{0.934} $\pm$ \textbf{0.002} \\ \bottomrule
\end{tabular}
\label{tab:results_manipulation_age}
\end{table*}

To fully validate our model, it is also necessary to understand how well it incorporates the information that is fed into the decoder and, consequently, how well it can manipulate attribute information within the speaker embedding.

To do so, we performed a set of experiments using the models trained for each attribute, where pre-trained classifiers are tested with regard to the "fake" attribute labels fed to the model's decoder.
Differently from the prior experiments, here, the "fake" attribute is random for every sample, as we want to cover both classes, for sex, and a widespread range of values for age. 
Specifically, we generate random logits using a distribution trained over the output logits of the external classifier for the training set. In the case of the sex classification model, to obtain the label of each vector of logits, we take the argmax and use the corresponding index.

We also test ASV performance, wherein the same information is used to condition both samples in same-speaker trials. For different speaker trials, different attribute information is used for either sample.

The results for this experiment are presented in Tables \ref{tab:results_manipulation} and \ref{tab:results_manipulation_age}. We do not report here Zebra metrics, as they measure information disclosure and, thus, are not relevant for this task.

Contrary to prior experiments, in this experiment, for sex information the full model does not clearly improve in terms of classification metrics over the adversarial loss-only model, with only small differences observed for the AUPRC (higher for the full model) and UAR (higher for the adversarial-only model). Nevertheless, in terms of ASV performance, the full model outperforms all models.

In the case of the age manipulation experiments, in Table \ref{tab:results_manipulation_age}, we observe a similar pattern, with the full and adversarial-only models showing only slight differences for CCC (higher for the adversarial-only model) and PCC (higher for the full model). For age, we also observe that the values obtained in terms of CCC and PCC are much higher (and improvement of $\sim$0.2) than those obtained for the original data, as opposed to what was shown by the sex information manipulation experiments, where the classification metrics presented some degradation when compared to the original data. We hypothesise that, in the case of sex information, some logit configurations may be very close to the classification boundary between the two classes, whereas for age, given that it is a regression task, this may happen less often.

The fact that the best models are able to achieve a 90\% UAR and 0.91 CCC for "fake" attribute prediction with pre-trained classifiers shows that our model is capable of manipulating the attribute information within the speaker embedding.
Moreover, the performance in terms of speaker verification is better than the performance obtained for the original experiments (cf. results in Tables \ref{tab:results_bat} and \ref{tab:results_age}), presenting a degradation of only 0.15\% EER when compared to the original data, for the sex manipulation model, and a $\sim$3\% EER degradation for the age manipulation model.
The likely reason for this is that the same attribute information is being used for same-speaker trials, and different information is being used for different-speaker trials.
In other words, embeddings corresponding to the same speaker will be transformed with the same "fake" information (i.e., the same random logits), bringing them closer together. 
Conversely, pairs of different speakers will be further apart, as the random logits will be different for each vector. This will make the pairs more discriminative and hence improve the speaker verification results.

\subsection{Cross-domain results}

\begin{figure*}[t]
\begin{subfigure}[a]{0.49\textwidth}
    \centering
    \includegraphics[width=0.5\textwidth]{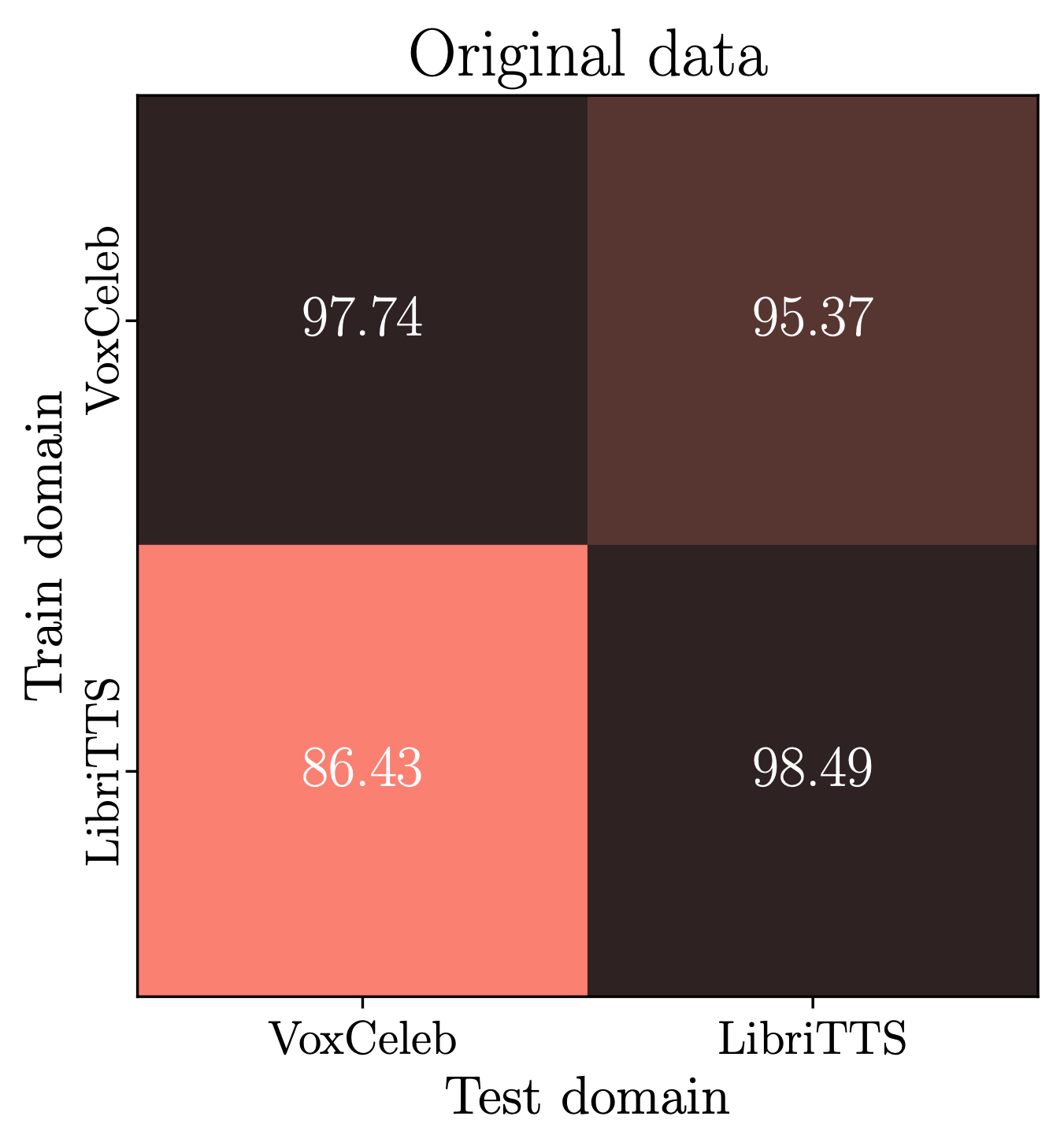}
    \captionsetup{justification=centering}
    \caption{Original data.}
    \label{fig:org_ood}
\end{subfigure}
\begin{subfigure}[a]{0.51\textwidth}
    \centering
    \includegraphics[width=\textwidth]{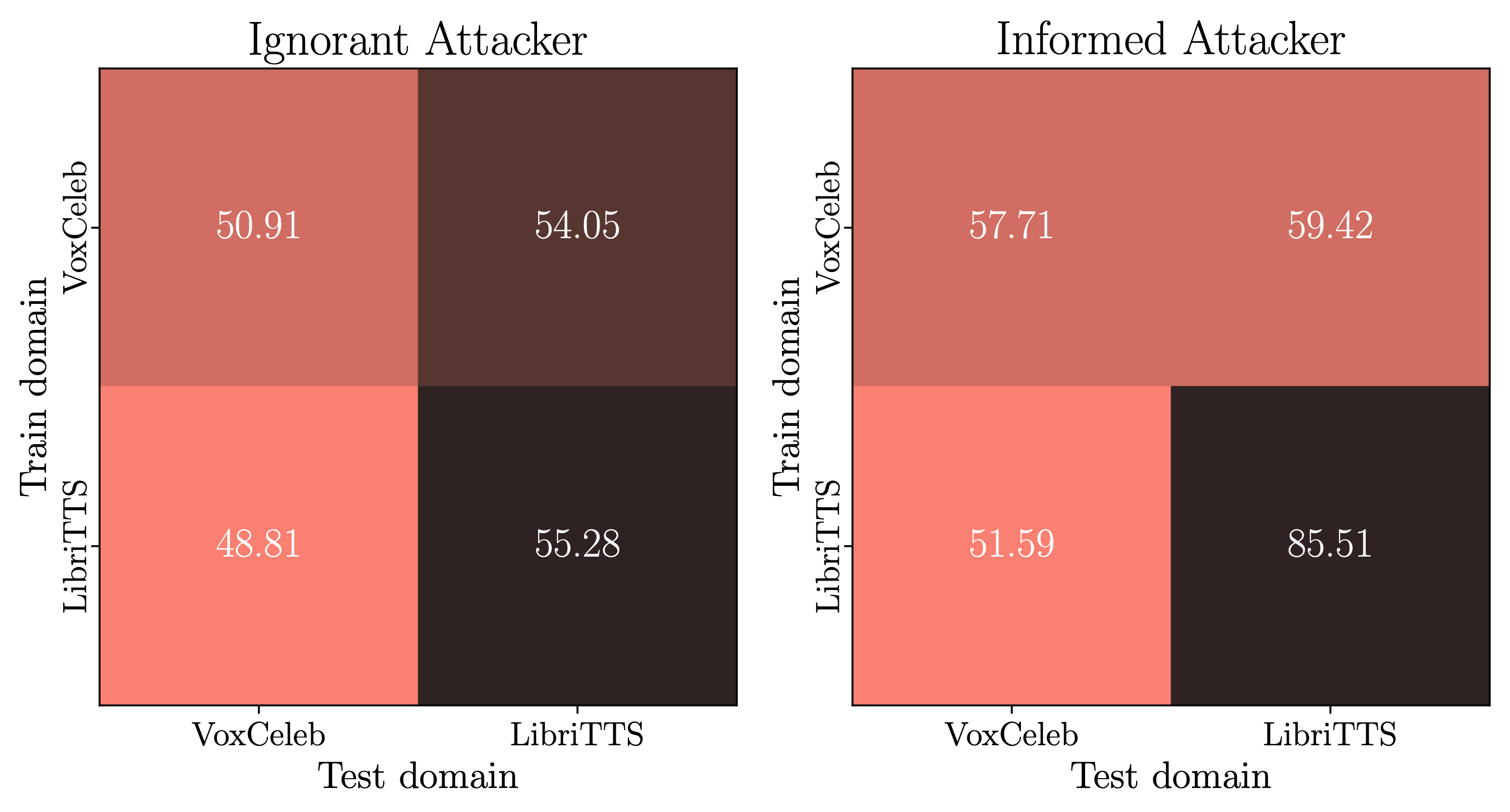}
    \captionsetup{justification=centering}
    \caption{VQ-VAE trained on VoxCeleb.}
    \label{fig:vox_ood}
\end{subfigure}

\hfill

\begin{subfigure}[b]{0.5\textwidth}
    \centering
    \includegraphics[width=\textwidth]{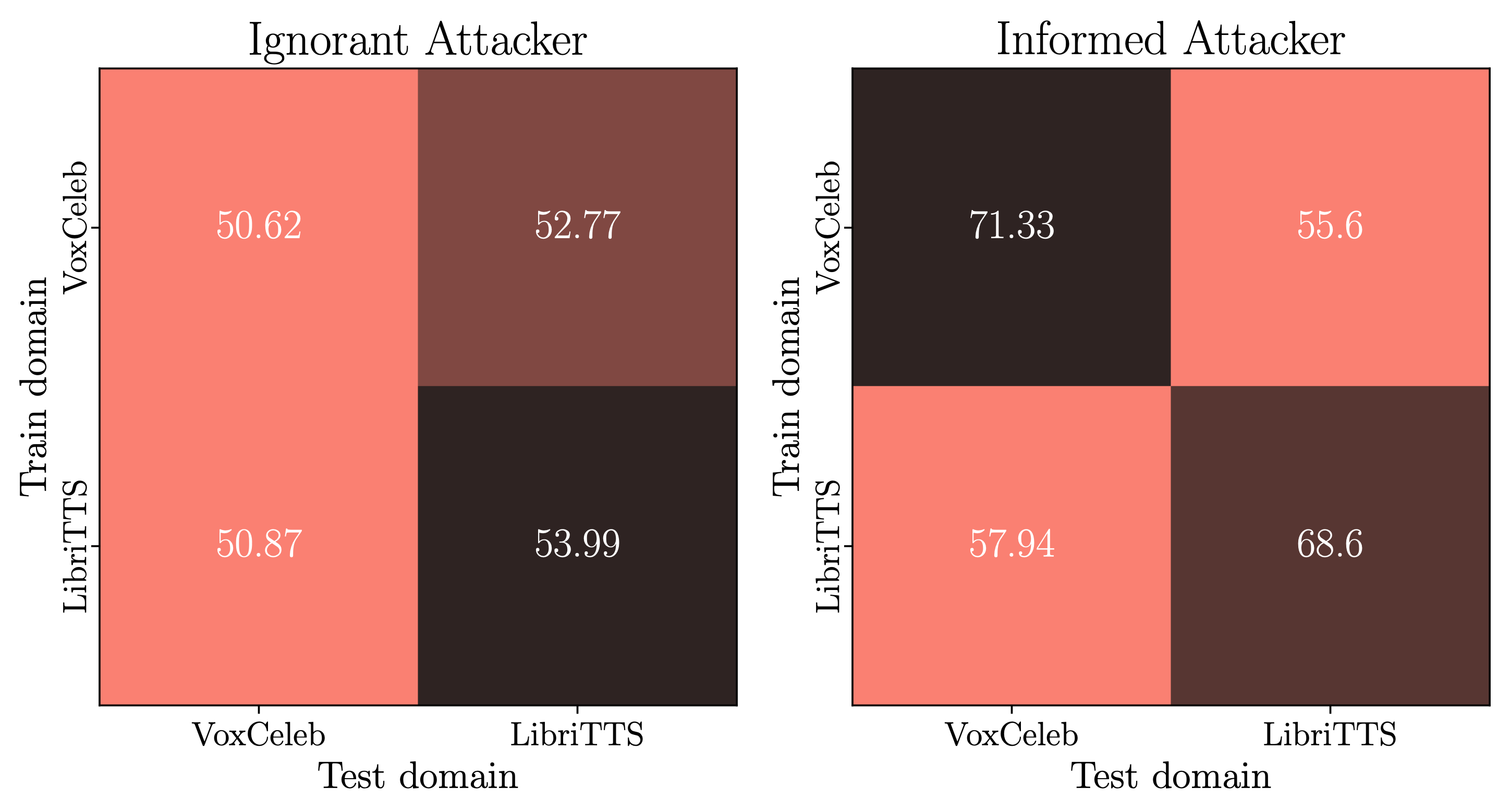}
    \captionsetup{justification=centering}
    \caption{VQ-VAE trained on LibriTTS.}
    \label{fig:libri_ood}
\end{subfigure}
\begin{subfigure}[b]{0.5\textwidth}
    \centering
    \includegraphics[width=\textwidth]{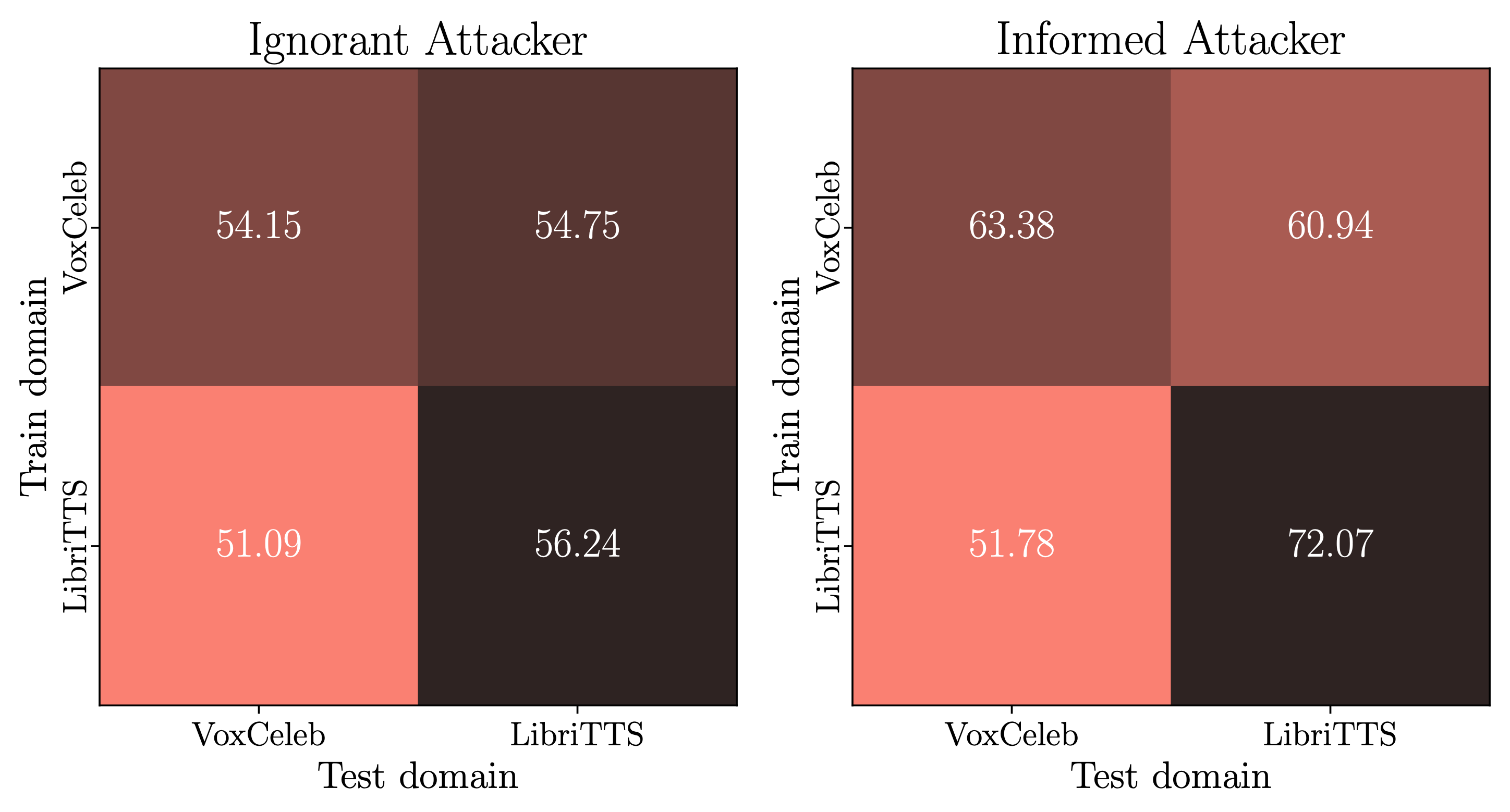}
    \captionsetup{justification=centering}
    \caption{VQ-VAE trained on 50\% VoxCeleb and 50\% LibriTTS.}
    \label{fig:vl_ood}
\end{subfigure}
\hfill
\captionsetup{justification=centering}
\caption{Results for the cross-dataset experiments.}
\label{fig:crossdataset}
\end{figure*}

In this section, we discuss the cross-domain experiments for the sex attribute.
These experiments aim to provide an understanding of how well our models can generalise their ability to remove attributes to unseen domains.
As stated in Section \ref{sec:eval}, we perform a total of 8 experiments (cf. Table~\ref{tab:crossdomain}), using two datasets (VoxCeleb and LibriTTS) to train the VQ-VAE and to train and test the attribute classifier.
These experiments are performed with the two types of attackers, ignorant and informed, as well as for the original non-manipulated data.
In total, this results in 28 experiments, the results of which can be found in Fig.~\ref{fig:crossdataset}. For conciseness, this figure only reports results in terms of mean UAR.
For every sub-figure, the Y-axis corresponds to the domain used to train the attribute classifier, whereas the X-axis corresponds to the domain of the test data. Darker colours indicate higher UAR values, and conversely, lighter colours indicate lower UAR values.

Regarding the cross-domain results for the original data, shown in Fig.~\ref{fig:org_ood}, we can observe that each domain tested against itself (diagonal squares) provides very high results, with the highest UAR for sex classification corresponding to attribute classifiers trained and tested on LibriTTS.
In the values in the counter-diagonal, whereas the classifier trained on VoxCeleb and tested on LibriTTS provides good results, around 95\% UAR, the opposite shows a UAR of around 86.5\%, amounting to an absolute degradation of almost 10\%.
This trend is observed in most of the remaining experiments, showing that sex attribute classifiers trained on LibriTTS do not generalise well to VoxCeleb. A possible reason for this is the fact that LibriTTS contains samples of read speech under very controlled conditions (Audiobooks), whereas VoxCeleb is composed of interviews recorded in very diverse and noisy conditions, making it easy for the classifier trained on VoxCeleb to obtain good results in the clean conditions of LibriTTS, and the opposite much harder.

For the data manipulated using the VQ-VAE model trained on VoxCeleb, in Fig.~\ref{fig:vox_ood}, we observe the same effects of training the attribute classifier on LibriTTS and testing it on VoxCeleb. 
However, considering the LibriTTS test results, we can see that our model is not able to perform as well as for VoxCeleb for both attackers. This is most evident for the informed attacker, where the sex classifier trained and tested on LibriTTS achieves an 85\% UAR, showing that the model is somewhat domain-specific.

To understand the source of the domain dependence in our method, we trained a VQ-VAE with LibriTTS and performed the same cross-domain experiments.
In Fig.~\ref{fig:libri_ood}, we see that the performance for the attribute classifier trained and tested on LibriTTS is much better for privacy, dropping around $17\%$ UAR, for the informed attacker, when compared to the VQ-VAE trained with VoxCeleb. 
Moreover, for the informed attacker, we observe almost equal performance when training and testing the attribute classifiers on the same domain or in cross-domain settings.
Nonetheless, the performance of the VQ-VAE for LibriTTS in the informed attacker scenario is not on par with the model trained on VoxCeleb. 
One of the reasons may be the fact that the model was trained with much less data: $\sim$205,000 utterances for LibriTTS versus $\sim$880,000 utterances for VoxCeleb.

Finally, we also explore the behaviour of our model when trained on both domains. To do so, we use the same amount of data taken for both datasets. 
In this case, we observe a degradation of the results when testing in the original VQ-VAE training domain.
However, when the model is tested across training domains (e.g., the VQ-VAE is trained on VoxCeleb and tested for privacy on LibriTTS), it performs better than the VQ-VAEs trained for individual domains.

Specifically, in the scenario where the attribute classifier was trained and tested on VoxCeleb, the result for the informed attacker presented in Fig.~\ref{fig:vl_ood}, shows a degradation of $\sim$5.5\% UAR when compared to the in-domain value presented in Fig.~\ref{fig:vox_ood}. 
Moreover, when considering the attribute classifiers trained and tested on LibriTTS, the result shown in Fig.~\ref{fig:vl_ood} presents a degradation of $\sim3.5$\% UAR, when compared to the in-domain result of Fig.~\ref{fig:libri_ood}. 
Contrarily, the attribute classifier trained and tested on LibriTTS, obtained using the out-of-domain VQ-VAE trained on VoxCeleb (cf. Fig.~\ref{fig:vox_ood}), the model trained on both datasets shows an improvement of $\sim$13\% UAR. In addition, the attribute classifier trained and tested on VoxCeleb shows an improvement of $\sim$8\% UAR, when compared to the out-of-domain VQ-VAE trained on LibriTTS (cf. Fig.~\ref{fig:libri_ood}).

This supports the argument that combining multiple domains in the training data helps increase the robustness of the model to those domains.

For the ignorant attacker, the performance is stable across the three experiments, with the results obtained for the VoxCeleb test set being close to chance level, and for the LibriTTS test set averaging around 54.5\%.

Overall, the results of these experiments for the informed attacker indicate that the performance of the VQ-VAEs is dependent on the domain of data they were trained on. On the other hand, for the ignorant attacker, the models' performance appears to be independent of the data used to train and test the attribute classifiers. Moreover, the general approach in itself seems to be independent, with our results showing that different models can be trained on data from specific domains to obtain better results in these domains.

\subsection{Limitations}

The results detailed in the previous sections show that the proposed method fulfils the objectives set at the end of Section \ref{sec:prelim}. 
Specifically, the trained models allow the suppression of the two target attributes, sex and age, achieving privacy results close to chance level in in-domain settings, as well as in several cross-domain settings.
Moreover, our experiments regarding sex information have shown that the proposed method is in fact able to manipulate attribute information, instead of simply removing it.

Nevertheless, the proposed method still presents some limitations. For instance, the sex and age attribute classification results show that our method is still unable to remove all attribute information. This means that, for stronger attackers, it may still be possible to recover this information.
On the other hand, the measure of the utility of the proposed method rests solely on ASV performance. To fully understand the impact of the proposed method, it would be important to evaluate its effects on the detection of other speaker traits or conditions which may be important for other downstream tasks.

In addition, the proposed method does not provide a clear way to trade off utility and privacy. For instance, the results pertaining to the age attribute that are shown in Table \ref{tab:results_age} indicate that as each component of the method is added, the speaker verification results degrade, whereas privacy improves. However, for sex information, this is not the case, and only the baseline VQ-VAE is able to achieve a better ASV result when compared to the full method (VQ-VAE + ADV + MI).
One could also consider changing the weights of each loss to manipulate this trade-off. However, our preliminary experiments -- wherein the weights for each loss were varied logarithmically between 0.1 and 1000 -- showed that this relation was not linear, i.e., increasing the losses' weights did not always correlate with either more privacy or less utility. We consider that making this trade-off clearer and easier to control is an important objective for future study.

\section{Conclusions}
\label{sec:conclusions}

%%%%%%%%%% Conclusions %%%%%%%%%%%%

In this work, we propose the use of a combination of a VQ-VAE, an adversarial classifier, and a Mutual Information loss to remove or manipulate sex and age information in speaker representations. 
Our model was tested in an Automatic Speaker Verification setting, where both the speaker representation extraction step and the application of our model are assumed to be performed in the user's device. Our model is much smaller ($\sim$1M parameters) than the speaker representation extraction model ($\sim$14M parameters), corresponding to a small additive cost.

The experiments that were conducted prove the validity of the proposed method and show that our model is able to drop the classification or estimation performance of both attributes to close to chance level while keeping the utility of the speaker representations for Automatic Speaker Verification. The proposed models were also successfully validated with regard to the manipulation of both attributes, and a cross-domain study further showed that our method still works when trained and tested with out-of-domain data.

The avenues for future work are vast, with numerous topics worth exploring.
In terms of privacy, the proposed method could be tested for the removal of multi-class attributes such as accent information. Other paralinguistic traits, such as emotional information could also be worth exploring.
Another possible extension of this work would be its application to domain generalisation, i.e., minimising the amount of domain information contained in speaker representations~\cite{li2023mutual}.
Alternatively, one could also explore the cross-attribute effect of each of the attribute models, for instance, by measuring the effect of the age removal model on sex classification performance and vice versa. This would allow a more in-depth understanding of the effects of attribute removal models. 
A similar line of work would be the application of each of the models in sequence to understand whether it is possible to remove both age and sex information from the same speaker embedding with the proposed methods.
Another potentially relevant research line would be the use of the proposed model in voice conversion and text-to-speech tasks, as a way to manipulate and control speaker traits, as well as to anonymise speech to some extent~\cite{noe2023hiding}. Training our model for these tasks would also show its applicability to different speaker representation extractors, as well as its robustness to different downstream applications.

The development of methods that hide speaker attributes raises the question of which attributes are more related to speaker identity, or which can considered more sensitive. One could ask if hiding age provides more privacy than hiding the speaker's sex, or if it would be more important to hide other speaker traits.
In a real-world scenario, it would be important to inform the user of not only the utility degradation introduced by the removal of certain attributes but also of the possible privacy protections that can be achieved by hiding each specific attribute. 
The fact that, in this work, we successfully test our approach for two attributes provides an indication of the generalisation capabilities of the method to any other attribute, and motivates the study of the removal of other attributes.

\section*{Appendix A}

In this Appendix, we describe the two Mutual Information (MI) estimators used in this work: (1) the 
Kraskov, St{\"{o}}gbauer and Grassberger (KSG) \cite{ksg, gao2018demystifying} estimator to estimate the MI between two continuous random variables; (2) the MI estimator proposed by B. Ross \cite{ross2014mutual}, for mixtures of discrete and continuous random variables. 
The descriptions contained in this Appendix closely follow the method descriptions presented in \cite{ksg,ross2014mutual}.

\subsubsection{Mutual information estimator for continuous random variables}

We will start by providing a high-level description of the continuous-continuous KSG mutual information estimator~\cite{ksg} and the intuition behind this estimator. Although it is only used for the manipulation of a continuous attribute (i.e., age), understanding this estimator will allow the reader to understand the intuition behind nearest-neighbour MI estimators and consequently understand the continuous-discrete MI estimator proposed by B. Ross \cite{ross2014mutual}.

The mutual information $I(Z, Y)$ between two continuous variables $Z$ and $Y$ can be expressed in terms of the individual differential entropies and the entropy between the two random variables:

\begin{equation}
\label{apdx:eq:mi_entropy}
    I(Z, Y) = H(Z) + H(Y) - H(Z, Y),
\end{equation}

\noindent having each $H(\cdot)$ defined as:
\begin{equation}
    H(S) = E[-\text{log}\,\mu_s(s)] = -\frac{1}{N}\sum_{i=1}^{N} \text{log}\, \mu_{s}(s_i),
    \label{apdx:eq:entropy_expected}
\end{equation}
where $S$ is any random variable and $\mu_s$ is its corresponding the probability density function.

Given a set of $N$ observations taken from dataset $\mathcal{D}$ of the joint variable $M = (Z, Y)$, $m_i = (z_i, y_i)$, with $i \in 1\, ...\, N$, the goal of an MI estimator is to use these observations to obtain $I(Z, Y)$. 

From eq. \eqref{apdx:eq:mi_entropy}, it is possible to see that the MI can be computed through its entropy terms.
However, it is not possible to  compute these terms directly because $\mu_{z}(z)$, $\mu_{y}(y)$ and $\mu_{z,y}(z,y)$) are unknown. 
Instead, one needs to leverage the observations and use them to estimate the value of each entropy term.

To do so, KSG applies the Kozachenko-Leonenko (KL)~\cite{kle} $k$-nearest neighbour entropy estimator.
This estimator works by defining a probability distribution $P_k(\epsilon)$ of the distance ($\epsilon/2$) between each sample $s_i$ -- sampled from a continuous random variable $S$ -- and its $k^{th}$ neighbour.

Let us consider that each $p_i$ corresponds to the mass of a $d_S$-dimensional $\epsilon$-ball around $s_i$, where $d_S$ is the dimensionality of $S$.
The KL estimator leverages the fact that, by estimating $p_i(\epsilon)$, it is possible to indirectly estimate the density $\mu_s(s_i)$ (assuming it is constant within the entire $\epsilon$-ball), since, by definition:
\begin{equation}
    \mu_s(s_i) \approx \frac{p_i(\epsilon)}{v_{d_s} \epsilon^{d_s}}
    \label{apdx:eq:density}
\end{equation}
where $v_{d_s}$ is the volume of the $d_S$-dimensional unit ball, and $\epsilon$ its radius. $v_{d_S} = 1$ for the maximum norm, and $v_{d_S} = \pi^{\frac{d_S}{2}}/\Gamma(\frac{d_S}{2} + 1)$ for the $l_2$ norm, with $\Gamma(\cdot)$ corresponding to the \textit{gamma} function.

Considering that $\epsilon_i^d$ can be computed for each sample $s_i$ -- it corresponds to twice the distance between $s_i$ and its $k^{th}$ neighbour -- to obtain the density it is only necessary to further compute $p_i(\epsilon)$. 
However, what is required is the expected value of $\mu_s(s_i)$. For this reason, in KL the expected value of $\mbox{log}(p_i)$ is computed directly~\cite{kle, ksg}:
\begin{equation}
    E[\mbox{log}\,(p_i)] = \psi(k) - \psi(N)
    \label{apdx:eq:p_expected}
\end{equation}
with $k$ being the pre-defined number of neighbours, $N$ the number of observations, and $\psi(\cdot)$ the \textit{digamma} function~\cite{abramowitz1988handbook}.

Combining eqs. \eqref{apdx:eq:entropy_expected}, \eqref{apdx:eq:density} and \eqref{apdx:eq:p_expected}, one obtains the full KL estimator:
\begin{equation}
    \hat{H}(S) = \psi(N) - \psi(k) + \text{log}\,(v_{d_s}) + \frac{d_s}{N} \sum_{i=1}^{N} \text{log}\,(\epsilon_i)
    \label{apdx:eq:hx}
\end{equation}
This can be extended to the joint random variable $M=(Z,Y)$, as:
\begin{equation}
\begin{split}
    \hat{H}(X, Y) & = \psi(N) - \psi(k) \\ 
    & + \text{log}\,(v_{d_Z}v_{d_Y}) + \frac{d_{Z}+d_{Y}}{N} \sum_{i=1}^{N} \text{log}\,\epsilon_i,
\end{split}
\label{apdx:eq:hxy}
\end{equation}

where $v_{d_Z}$ and $v_{d_Y}$ correspond to the volume of the $d_Z$ and $d_Y$-dimensional unit balls and $\epsilon_i/2$ corresponds to the distance between two observations in the joint space $Z$.

To obtain $I(Z, Y)$ one could simply apply eqs. \eqref{apdx:eq:hx} and \eqref{apdx:eq:hxy}. However, the distance scales of the joint space $Z$, and variables $Z$ and $Y$ may be very different. 
To circumvent this issue, the KSG estimator (specifically, Algorithm (2) of \cite{ksg}) first finds the $k^{th}$ neighbour of sample $m_i$ in the joint space $M$, with distance $\epsilon_i/2$, using the maximum norm $\norm{m - m'} = \text{max}\{\norm{z - z'}, \norm{y - y'}\}$, for any metric space in $X$ or $Y$. 
It then considers the number of points $n_{s_{i}}$ that are within distance $\epsilon_{s_i}/2$ for each of the marginal sub-spaces of $Z$ and $Y$, as a replacement of the original fixed number of neighbours $k$. This yields a second estimator $\hat{H}(S)$ for the differential entropies:
\begin{equation}
\begin{split}
    \hat{H}(S) &= \psi(N) - \frac{1}{N} \sum_{i=1}^{N} \psi(n_{s_{i}} + 1) \\ 
    & - \text{log}\,(v_{d_S}) - \frac{d_S}{N}\sum_{i=1}^{N}\text{log}\, \epsilon_{s_{i}},
\end{split}
\label{apdx:eq:hx2}
\end{equation}

where $S$ corresponds to either $Z$ or $Y$.
Finally, by combining equations \eqref{apdx:eq:hxy} and \eqref{apdx:eq:hx2}, results in:

\begin{equation}
    \hat{I}(Z, Y) = \psi(k) + \psi(N) - \langle \psi(n_z + 1) + \psi(n_y + 1) \rangle,
\end{equation}

\noindent where $\langle ... \rangle = \frac{1}{N}\sum_{i=1}^{N} ... $ is the average operator.

In our preliminary experiments, we found that this estimator was not able to perform well when large differences in the dimensionality of each marginal space occurred, or when very different scales of $X$ and $Y$ were present, a result that is consistent with what is reported in the literature \cite{gao2015efficient}. Instead, we used the adapted estimator of Gao et al. \cite{gao2018demystifying}, which introduces a bias-correction term that accounts for the volumes in each dimension, and that uses the $l_2$ distance instead of the maximum norm~\cite{gao2018demystifying}:
\begin{equation}
\begin{split}
    \hat{I}(Z, Y) & = \text{log}(N) + \psi(k) + \text{log}\frac{v_{z} v_{y}}{v_{z}+v_{y}} \\
    & - \langle \text{log}(n_z) + \text{log}(n_y) \rangle,
\end{split}
\label{apdx:eq:mi_cc}
\end{equation}

\subsubsection{Mutual information estimator for discrete and continuous random variables}

The continuous-discrete MI estimator proposed by Ross \cite{ross2014mutual} applies a similar idea to that of Kraskov et al. \cite{ksg}, leveraging the k-nearest neighbour KL entropy estimator~\cite{kle}.

From eq. \eqref{apdx:eq:mi_entropy}, it can be shown that for a discrete random variable~$Y$, and a continuous random variable~$Z$~\cite{ross2014mutual}:
\begin{equation}
    I(X, Y) = - \langle \,\text{log}\,\mu_{z}(z) \rangle + \langle \, \text{log}\,\mu_{z|y}(z|y) \rangle.
\end{equation}

Using this, the author then applies the KL differential entropy estimator (cf. eq. \eqref{apdx:eq:hx}) twice, to estimate each term. This leads to:
\begin{equation}
    \hat{I}(z_i, y_i) = \psi(N) + \psi(k) - \psi(N_{y_i}) - \psi(n_{z_i}),
    \label{apdx:eq:mi_xvec}
\end{equation}

where $I(z_i, y_i)$ is the mutual information for a single observation $(z_i, y_i)$, and where $N_{y_i}$ corresponds to number of samples in $\mathcal{D}$ with the same discrete value $y_i$. This is relevant as it shows that the notion of neighbour changes from the previous estimator, and instead a sample is only considered a "neighbour" if it comes from the subset of $\mathcal{D}$ where $Y\!=\!y_i$.
For this reason, $\epsilon/2$ is set as the distance between $z_i$ and the $k^{th}$ sample that shares the same value $y_i$, and $n_{z_i}$ is counted as the number of samples, now for the full set of $\mathcal{D}$, that are within this distance.

Finally, to compute the MI for the full set of samples, one computes the average of all $I_i(z_i, y_i)$:

\begin{equation}
    \hat{I}(X, Y) = \psi(N) + \psi(k) - \langle \psi(N_y)\rangle - \langle \psi(n_{z}) \rangle.
    \label{apdx:eq:mi_cd}
\end{equation}

\section*{Acknowledgement}
We thank the authors of \cite{noe2022bridge} for their help in reproducing their work. 

\bibliographystyle{unsrt}
\bibliography{mybib}

\begin{IEEEbiography}[{\includegraphics[width=1in,height=1.25in,clip,keepaspectratio]{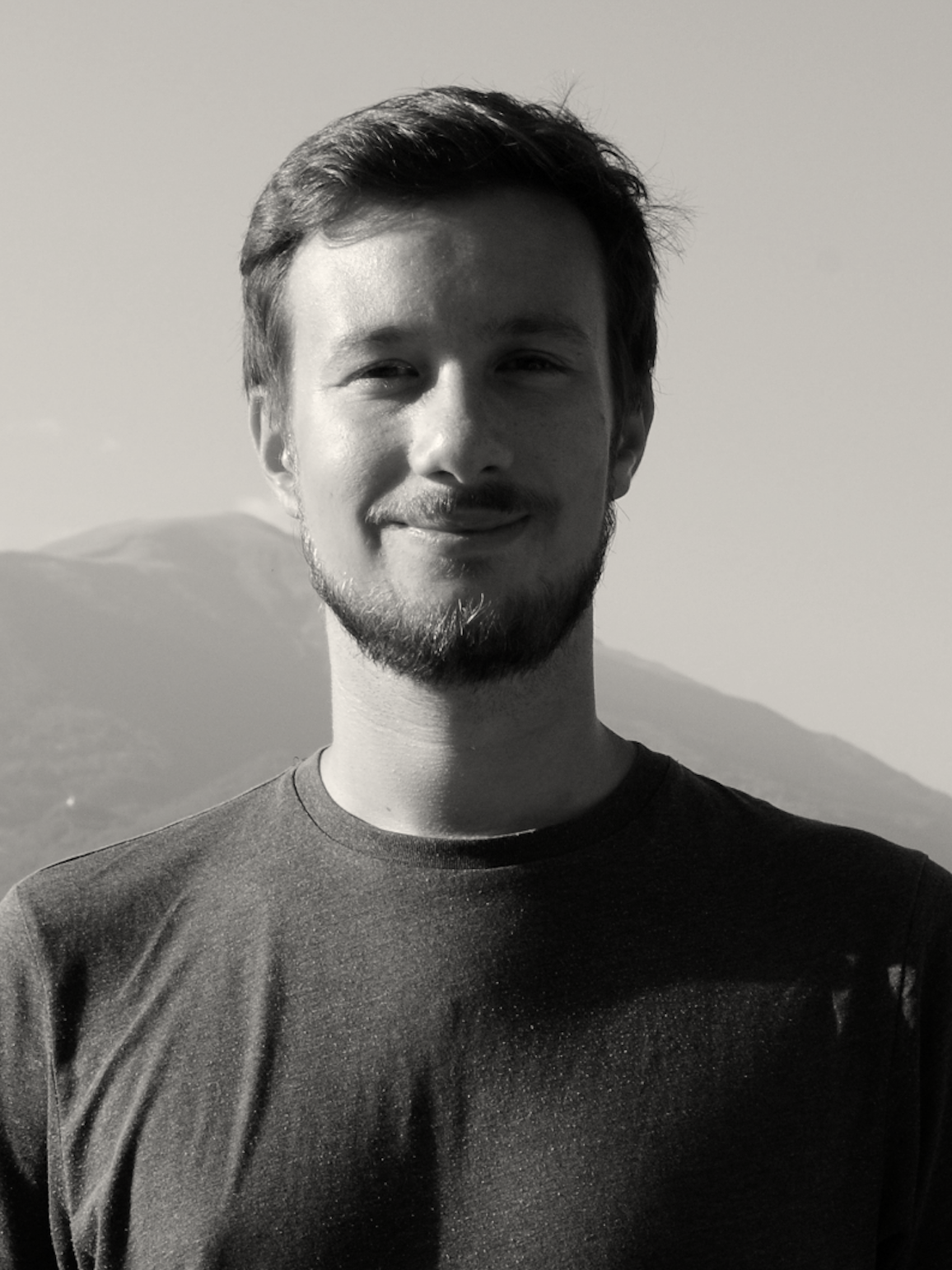}}]{Francisco Teixeira}
received the B.S. and M.Sc. degrees in Electrical and Computer Engineering from Universidade de Lisboa, Lisbon, Portugal in 2018 and is currently pursuing a Ph.D. degree in Electrical and Computer Engineering at the same university.
His main research interest is privacy-preserving speech processing. In particular, his research focuses on the combination of cryptographic and machine learning methods for privacy in remote speech processing settings.
He is a member of the International Speech Communication Association - Student Advisory Committee (ISCA-SAC).
\end{IEEEbiography}

\begin{IEEEbiography}[{\includegraphics[width=1in,height=1.25in,clip,keepaspectratio]{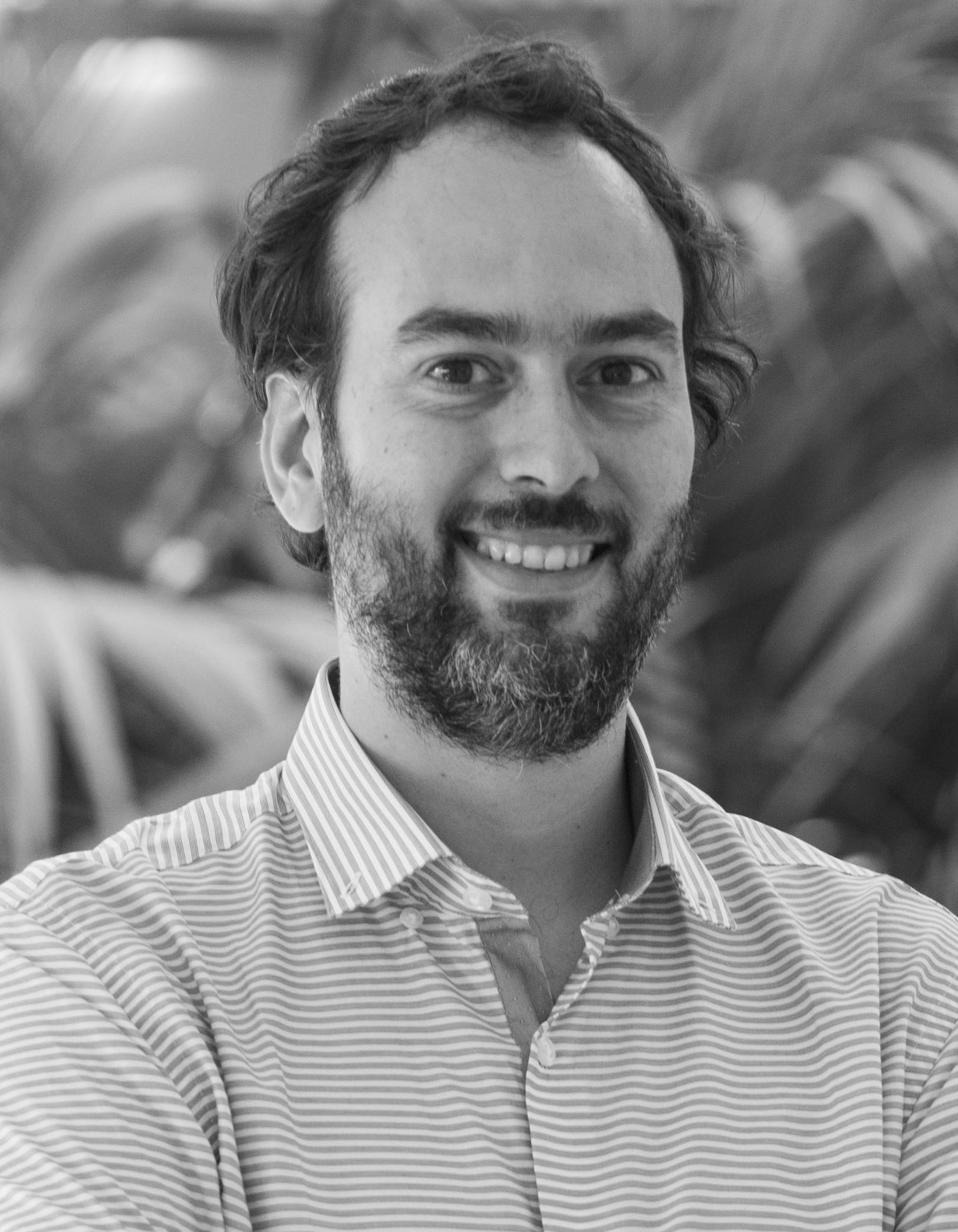}}]{Alberto Abad}
received the Telecommunication Engineering degree from the Technical University of Catalonia (UPC), Barcelona, Spain, in 2002 and the Ph.D. degree from UPC, in 2007. Currently, he is an Associate Professor at the Department of Computer Science and Engineering (DEI) of Instituto Superior Técnico (IST) and a researcher at INESC-ID. He is the coordinator of the Human Language Technologies laboratory at INESC-ID and the deputy coordinator of the Master in Computer Science and Engineering of IST. He is also an IEEE Senior member.
His research interests include robust speech recognition, speaker and language characterisation, applied machine learning, healthcare applications, and privacy-preserving speech processing and machine learning. 
\end{IEEEbiography}

\begin{IEEEbiography}[{\includegraphics[width=1in,height=1.25in,clip,keepaspectratio]{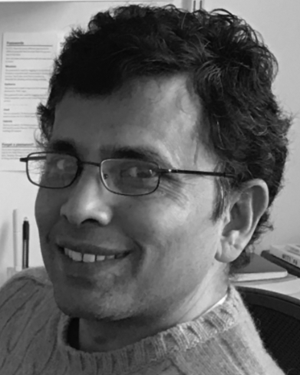}}]{Bhiksha Raj} 
Bhiksha Raj, IEEE Fellow, received a PhD degree in electrical and computer engineering from Carnegie Mellon University, Pittsburgh, PA, USA, in 2000. He is currently a professor in the Computer Science Department, at Carnegie Mellon University where he leads the Machine Learning for Signal Processing Group. He joined the Carnegie Mellon faculty in 2009, after spending time with the Compaq Cambridge Research Labs and Mitsubishi Electric Research Labs. He has devoted his career to developing speech and
audio-processing technology. He has had several seminal contributions in the areas of robust speech recognition, audio content analysis and signal enhancement, and has pioneered the area of privacy-preserving speech processing. He is also the chief architect of the popular Sphinx-4 speech-recognition system.
\end{IEEEbiography}

\begin{IEEEbiography}[{\includegraphics[width=1in,height=1.25in,clip,keepaspectratio]{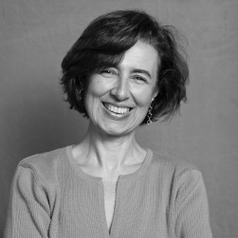}}]{Isabel Trancoso} is a former full professor at IST (Univ. Lisbon) and President of the Scientific Council of INESC-ID. She got her PhD in ECE from IST in 1987. She chaired the ECE Department of IST. She was Editor-in-Chief of the IEEE Transactions on Speech and Audio Processing and had many leadership roles in SPS (Signal Processing Society of IEEE) and ISCA (International Speech Communication Association), namely having been President of ISCA and Chair of the Fellow Evaluation Committees of both SPS and ISCA. Although recently retired, she is still actively supervising students and playing relevant roles in professional associations, such as Vice-Chair and Chair of the IEEE Fellow Committee (2023, 2024). She was elevated to IEEE Fellow in 2011, and to ISCA Fellow in 2014.
\end{IEEEbiography}

\EOD

\end{document}